\documentclass[english,onecolumn,draftcls]{IEEEtran}
\usepackage[T1]{fontenc}
\usepackage[latin9]{inputenc}
\usepackage{color}
\usepackage{amsmath}
\usepackage{amsthm}
\usepackage{amssymb}
\usepackage{graphicx}

\makeatletter
\theoremstyle{plain}
\newtheorem{thm}{\protect\theoremname}
\theoremstyle{plain}
\newtheorem{prop}[thm]{\protect\propositionname}
\theoremstyle{plain}
\newtheorem{cor}[thm]{\protect\corollaryname}

\makeatother

\usepackage{babel}
\usepackage{cite}

\providecommand{\corollaryname}{Corollary}
\providecommand{\propositionname}{Proposition}
\providecommand{\theoremname}{Theorem}

\begin{document}
\title{Set-membership target search and tracking\\
 with cooperating UAVs using vision systems\footnote{This work has been submitted to the Elsevier / ScienceDirect for possible publication. Copyright may be transferred without notice, after which this version may no longer be accessible.}}
\author{Maxime Zagar$^{1,2}$, Luc Meyer$^{2}$, Michel Kieffer$^{1}$, H\'el\`ene Piet-Lahanier$^{2}$
\thanks{$^1$ Universit\'e Paris-Saclay, CentraleSup\'elec, CNRS, L2S, 91192 Gif-sur-Yvette, France {\tt\small first\_name.last\_name@l2s.centralesupelec.fr}}
\thanks{$^2$ DTIS, ONERA Universit\'e Paris-Saclay, Palaiseau, France {\tt\small first\_name.last\_name @ onera.fr}}}

\maketitle

\begin{abstract}
This paper addresses the problem of target search and tracking using
a fleet of cooperating UAVs evolving in some unknown region of interest
containing an \emph{a priori} unknown number of moving ground targets.
Each UAV is equipped with an embedded Computer Vision System (CVS),
providing an image with labeled pixels and a depth map of the observed
part of its environment. Moreover, when a target is identified, a
box containing the corresponding pixels in the image is also provided.
Hypotheses regarding information related to the classified pixels,
the depth map, and the boxes are introduced to allow its exploitation
by set-membership approaches. Using information provided by the CVS
and these hypotheses, each UAV evaluates sets guaranteed to contain
the location of the identified targets and a set containing the possible
locations of targets still to be identified. Then, each UAV uses these
sets to design its trajectory to search and track targets. The efficiency
of the proposed approach is illustrated via simulations.
\end{abstract}

\section{Introduction\label{sec:Introduction}}

The problem of searching and tracking targets using a fleet of Unmanned
Aerial Vehicles (UAVs) in some Region of Interest (RoI) has attracted
considerable attention in recent years, see, \emph{e.g.}, \cite{robin_multi-robot_2016,queralta_collaborative_2020,abood_survey_2022,lyu_unmanned_2023}
and the references therein. The difficulty of this problem depends
(\emph{i}) on the knowledge available on the environment and on whether
it is structured or not, (\emph{ii}) on the type and quality of information
provided by the sensors embedded in the UAVs, and (\emph{iii}) on
the \emph{a priori} knowledge about targets (shape, number, dynamics).

Cooperative Search, Acquisition, and Track (CSAT) problems in possibly
unknown and cluttered environments are still challenging \cite{lyu_unmanned_2023}.
In absence of prior knowledge, a representation of the environment
is usually constructed in parallel to the search for targets \cite{li_target_2021,ji_source_2022}
using techniques such as those presented in \cite{placed_survey_2023}.
For that purpose, UAVs must be equipped with a Computer Vision System
(CVS) including a camera and image processing algorithms. Many prior
works assume that a UAV gets a noisy measurement of the state of targets
present in its Field of View (FoV) \cite{kuhlman_multipass_2017,zhu_multi-uav_2021,zhao_distributed_2022,ibenthal_localization_2023,zhang_enhancing_2024}.
The complex processing performed by the CVS to get the state measurement
from data acquired by the camera is usually ignored. As a consequence,
the estimation uncertainty associated to the state measurements is
difficult to characterize, and is thus roughly approximated, which
is a major limit of those approaches.

This paper presents a CSAT approach for ground targets evolving in
a structured environment for which no prior map is available. UAVs
embed a CVS consisting of a camera with a limited FoV, a pixel classifier,
a depth map evaluation algorithm, and a target detection algorithm.
Hypotheses are introduced regarding the information obtained from
these algorithms to facilitate its processing by the UAVs. A distributed
set-membership estimation approach is proposed to exploit the CVS
information so as to obtain sets containing the locations of identified
targets compliant with the considered hypotheses. A set containing
all locations of targets still to be identified is also evaluated.
CVS information is also exploited to build sets where there is no
target. Each UAV exploits the information provided by its own CVS
and then accounts for the information broadcast by its neighbors.
The trajectory of each UAV of the fleet is then evaluated using a
variant of the Model Predictive Control (MPC) of \cite{ibenthal_bounded-error_2021}
so as to minimize the overall estimation uncertainty characterized
by the size of the set estimates.

To the best of our knowledge, this paper presents the first approach
to address a CSAT problem by a distributed set-membership approach
exploiting directly CVS information. Its main contributions lie (\emph{i})
in the introduction of hypotheses to exploit the information provided
by the CVS in a set-membership context, (\emph{ii}) in the proposition
of the associated set-membership estimation approach, and (\emph{iii})
in the consideration of an unknown structured environment where obstacles
may hide targets. The performance of the proposed approach is evaluated
via a simulation in a simplified urban environment.

Section~\ref{sec:Related-Works} reviews some related work. The CSAT
problem is formulated in Section~\ref{sec:Hypotheses} after a description
of models for the targets and the UAVs. Section~\ref{sec:HypCVS}
introduces models and hypotheses related to the information collected
by the CVS so that it may be exploited in a set-membership context.
The set-membership estimator is detailed in Section~\ref{sec:ExploitingCVS}
and in Section~\ref{sec:Prediction_Correction}. Section~\ref{sec:MPC}
briefly summarizes the distributed MPC approach to design the trajectory
of each UAV. Simulation results are provided in Section~\ref{sec:Simulations}.
Section~\ref{sec:Conclusion} concludes the paper and provides some
perspectives.

\section{Related Works\label{sec:Related-Works}}

CSAT approaches using a fleet of UAVs require finding, identifying,
locating, and tracking an unknown number of targets in some RoI \cite{queralta_collaborative_2020,lyu_unmanned_2023}.
This necessitates collecting measurements allowing to discriminate
the zones where a target can be located from the zones free of targets.
When a target has been identified and its location estimated, the
UAVs must keep track of its displacements while pursuing the exploration
of the RoI. Therefore, the trajectories of the UAVs result in a compromise
between exploratory search and target tracking. Many approaches have
been proposed to achieve parts of these various goals, fewer address
all of them, especially in an unknown environment.

\subsection{Representation of the RoI}

The simplest description of ROI is a cuboid or a simple geometric
shape without obstacle. In this context, search trajectories for a
fleet of UAVs have been designed in \cite{dames_distributed_2020,hou_uav_2023,banerjee_decentralized_2024}.
Nevertheless, the efficiency of these approaches decreases in presence
of occlusions due to obstacles, leading to possible non-detections.

To account for the presence of obstacles, the fleet of UAVs may either
exploit some \emph{a priori} known map of the environment or build
a map during the exploration. Exploration in cluttered environment
without any map has been considered in \cite{tang_novel_2019}, where
a swarm of robots is driven to a target emitting some signal while
using an obstacle collision avoidance approach. When several possibly
partly hidden targets have to be found, \cite{ibenthal_localization_2023}
considers groups of UAVs, each observing a part of the RoI with complementary
points of view. This limits the risk of a target being occluded from
all points of view and does not require building a map of the environment.
The price to be paid is an increase of the size of the fleet. Using
a known map, the UAV trajectories can be efficiently determined \cite{vanegas_uav_2016,goldhoorn_searching_2018,zhu_multi-uav_2021}.
Occlusions of the FoV by obstacles may be taken into account, as in
\cite{meera_obstacle-aware_2019,reboul_cooperative_2019}.

Many Simultaneous Localization and Mapping (SLAM) algorithms have
been developed to build maps, see the survey \cite{placed_survey_2023}.
The map may be represented by an occupancy grid \cite{hardouin_next-best-view_2020},
an OcTree \cite{asgharivaskasi_semantic_2023}, or a point cloud \cite{zhang_vehicle_2023}.
The availability of semantic information improves the mapping algorithms
\cite{atanasov_nonmyopic_2014,zhong_detect-slam_2018,zhang_fast_2022,asgharivaskasi_semantic_2023,serdel_smana_2023},
for example to distinguish static and mobile objects as in \cite{zhong_detect-slam_2018}.
In outdoor environment mapping, \cite{zhang_fast_2022,serdel_smana_2023}
use semantic information to evaluate the \emph{traversability} between
two locations, \emph{i.e.}, the possibility to reach one location
from another. Nevertheless, 3D mapping and exploration of large outdoor
environments is still challenging \cite{placed_survey_2023}, especially
for aerial vehicles due to their limited computational power. For
example, the mapping of a city-scale environment with an occupancy
grid raises storage issues. OcTree mapping may reduce this complexity
as shown in \cite{asgharivaskasi_semantic_2023}, but their study
is limited to a ground exploration of areas of $150\ \text{m}\times150\ \text{m}$.

CSAT algorithms including map construction during exploration have
been considered in \cite{li_target_2021,ji_source_2022}. These approaches
consider either a 2D \cite{ji_source_2022} or a 3D \cite{li_target_2021}
description of the RoI, but neglect potential occlusions of the FoV
by obstacles in the trajectory design.

\subsection{Environment perception and target detection}

Environment exploration, target detection, and map building heavily
rely on the information provided by the UAV sensors, such as images
provided by cameras or depth measurements obtained either from LiDARs,
or by processing the acquired images \cite{park_stereo_2012,jeon_accurate_2015,madhuanand_self-supervised_2021,shimada_fast_2023}.
Several computer vision algorithm may then be exploited such as image
or point cloud segmentation \cite{grilli_review_2017,howard_searching_2019,minaee_image_2021}.
Target detection techniques \cite{redmon_you_2016,minaeian_effective_2018,luo_yolod_2022,jiang_review_2022}
may then be put at work.

Once a target has been detected, several approaches have been proposed
to estimate its location. In \cite{minaeian_vision-based_2015,vanegas_uav_2016,sun_camera-based_2016,wang_real-time_2016,liu_novel_2018,liu_vision-aware_2022},
images of an RGB camera are used to estimate the location of a target
in a reference frame from its location in the images. These approaches
exploit a specific pixel, generally belonging to a 2D bounding box
containing the pixels associated to the target. The selection of the
pixel most representative of the target location is often heuristic
and may lead to relatively large localization errors, as evidenced
in \cite{di_gennaro_sensor_2023} when estimating the location of
humans evolving in the overlapping FoVs of static cameras. In \cite{kenmogne_cooperative_2019},
images of ArUco markers with known positions are acquired by UAVs,
processed by a CVS, and exploited by a set-membership approach to
cooperatively estimate the pose of the UAVs. Deep learning techniques
are now commonly used for target location estimation. For example,
YOLO \cite{redmon_you_2016} is used to localize a human in an unknown
environment present in the FoV of a UAV in \cite{ajmera_autonomous_2020}.
When the target remains in the FoV, it is tracked via optical-flow
techniques. In \cite{liu_target_2022}, YOLO exploits RGB images and
depth information to detect and localize ArUco markers placed on objects.
Using strong \textit{a priori} information such as 3D models of the
target shapes, deep learning algorithms exploiting images and depth
information, are able to estimate the location and orientation of
static objects \cite{atanasov_nonmyopic_2014,zeng_semantic_2018},
or vehicles \cite{sahin_review_2020,fan_deep_2022}. Nevertheless,
these algorithms may provide erroneous estimates with errors difficult
to characterize \cite{mousavian_3d_2017}.

\subsection{Estimates of a target location}

To represent the estimated target locations, a probability map is
often considered, which requires a a spatial discretization of the
RoI \cite{goldhoorn_searching_2018,zhen_intelligent_2020}. Nevertheless,
the performance of grid-based approaches heavily relies on the choice
of the cell dimension, and extension to large RoI is an issue \cite{placed_survey_2023}.
Alternative approaches such as \cite{dames_distributed_2020,banerjee_decentralized_2024}
combine a probability hypothesis density filter and random finite
sets. Assuming a bounded measurement noise, set-membership techniques
may be used as in \cite{ibenthal_bounded-error_2021} to get sets
guaranteed to contain the actual locations of targets, provided that
the hypotheses on the noise bounds are satisfied.

Determining parts of RoI free of targets may be very helpful in the
exploration process to reduce the estimation uncertainty of target
locations. The absence or presence of a target within an area monitored
by a UAV depends on the capacity of the embedded CVS to detect the
presence of a target \cite{allik_tracking_2019}, often represented
by a probability of detection. Nevertheless, this probability is highly
dependent on the observation conditions (UAV point of view, possible
occlusions) as evidenced in \cite{symington_probabilistic_2010} and
\cite{meera_obstacle-aware_2019}. As its characterization is difficult,
it is used in many approaches \cite{allik_tracking_2019,dames_distributed_2020,li_target_2021,hou_uav_2023},
but most often without justification. Alternatively, deterministic,
point-of-view dependent, target detection conditions have been introduced
in \cite{ibenthal_localization_2023}. A target may be detected and
identified only when it is observed from some set of point of views.
The set of point of views may be deduced from a map of the environment,
as in \cite{reboul_cooperative_2019}. Alternatively, the UAVs may
be organized to observe simultaneously parts of the RoI from a large
variety of point of views \cite{ibenthal_localization_2023}.

\subsection{Trade-off between exploration and tracking}

When mobile targets outnumber the UAVs, it proves difficult to maintian
constant monitoring of the detected targets. Different approaches
have been proposed to trade off environment exploration and target
tracking. Pre-planned trajectories are used in \cite{niedzielski_first_2021,yanmaz_joint_2023}.
In \cite{meera_obstacle-aware_2019}, several next best viewpoints
for each UAV are first identified using the approach of \cite{singh_modeling_2010,hardouin_next-best-view_2020}.
Then, polynomial trajectories are designed passing through these viewpoints
while avoiding known obstacles. A potential-field based path-planning
approach is developed in \cite{li_target_2021} to reach a location
in a 3D cluttered environment. The goal location is a cell characterized
by a higher uncertainty about the presence of a target than other
cells. In \cite{dames_distributed_2020}, each UAV updates its trajectory
with new weighted center of Voronoi cells obtained once the random
finite sets describing the possible target locations are updated.
Trajectories are designed in \cite{ajmera_autonomous_2020,hou_uav_2023,zhang_enhancing_2024}
via reinforcement learning techniques. A MPC approach is used in \cite{ibenthal_bounded-error_2021}
to design UAV trajectories minimizing the predicted estimation uncertainty
corresponding to the area of the set estimates related to identified
targets and of the set still to be explored.

\section{Models and problem formulation\label{sec:Hypotheses}}

Consider an urban environment to which a frame $\mathcal{F}$ is attached.
The RoI $\mathbb{X}_{0}\subset\mathbb{R}^{2}\times\mathbb{R}^{+}$
is a subset of the environment and its ground $\mathbb{X}_{\text{g}}$
is assumed to be flat, \textit{i.e.}, $\mathbb{X}_{\text{g}}=\left\{ \boldsymbol{x}\in\mathbb{X}_{0}\mid x_{3}=0\right\} $,
$x_{3}$ being the altitude of $\boldsymbol{x}$. $N^{\text{o}}$
static obstacles are spread within $\mathbb{X}_{0}$, with $N^{\text{o}}$
unknown. The shape $\mathbb{S}_{m}^{\text{o}}$ of the by the $m$-th
obstacle, $m\in\mathcal{N}^{\text{o}}=\left\{ 1,\dots,N^{\text{o}}\right\} $,
is the part of $\mathbb{X}_{0}$ it occupies. $\mathbb{S}_{m}^{\text{o}}$
is unknown but for all $m\in\mathcal{N}^{\text{o}}$, but we assume
that
\begin{equation}
\forall\boldsymbol{x}\in\mathbb{S}_{m}^{\text{o}},\forall\lambda\in\left[0,1\right],\lambda\boldsymbol{x}+\left(1-\lambda\right)\boldsymbol{p}_{\text{g}}\left(\boldsymbol{x}\right)\in\mathbb{S}_{m}^{\text{o}},\label{eq:Obs_Zconvex}
\end{equation}
where $\boldsymbol{p}_{\text{g}}\left(\boldsymbol{x}\right)$ is the
projection of $\boldsymbol{x}$ on the ground $\mathbb{X}_{\text{g}}$.
The assumption (\ref{eq:Obs_Zconvex}) is consistent with the representation
of buildings by a pyramidal stack of parallelipipeds introduced by
\cite{zhang_vehicle_2023}.

A fleet of $N^{\text{u}}$ UAVs with indexes in the set $\mathcal{N}^{\text{u}}=\left\{ 1,\dots,N^{\text{u}}\right\} $
is deployed in the RoI. The UAVs search a fixed but unknown number
$N^{\text{t}}$ of ground targets with indexes in the set $\mathcal{N}^{\text{t}}=\left\{ 1,\dots,N^{\text{t}}\right\} $.
We assume that the targets never leave $\mathbb{X}_{0}$ nor enter
in any obstacle.

Sections~\ref{subsec:Target-model} and~\ref{subsec:UAV-model}
introduce the models and assumptions considered for targets and UAVs.
Then Section~\ref{subsec:Problem-formulation} formalizes the CSAT
problem. In what follows, the time is sampled with a period $T$ and
$k$ refers to the time index. Several frames are introduced, such
as the frames attached to a UAV and to its camera. To lighten the
notations, vectors with no superscript related to a frame are implicitly
expressed in $\mathcal{F}$.

\subsection{Target model\label{subsec:Target-model}}

At time $t_{k}=kT$, the state of target $j\in\mathcal{N}^{\text{t}}$
is $\mathbf{x}_{j,k}^{\text{t}}$. The vector $\boldsymbol{x}_{j,k}^{\text{t}}$
containing the first three components of $\mathbf{x}_{j,k}^{\text{t}}$
gathers the coordinates of the center of gravity of target~$j$.
The projection $\boldsymbol{x}_{j,k}^{\text{t,g}}=\boldsymbol{p}_{\text{g}}\left(\boldsymbol{x}_{j,k}^{\text{t}}\right)$
of $\boldsymbol{x}_{j,k}^{\text{t}}$ on $\mathbb{X}_{\text{g}}$represents
the target location at time $t_{k}$. It is assumed to evolve as
\begin{equation}
\boldsymbol{x}_{j,k+1}^{\text{t,g}}=\mathbf{f}^{\text{t}}\left(\boldsymbol{x}_{j,k}^{\text{t,g}},\boldsymbol{v}_{j,k}^{\text{t}}\right),\label{eq:DynTarget}
\end{equation}
where $\mathbf{f}^{\text{t}}$ is known and $\boldsymbol{v}_{j,k}^{\text{t}}$
is some unknown target control input only assumed to belong to a known
box $\left[\boldsymbol{v}^{\text{t}}\right]$. The space occupied
by target~$j$ is the subset $\mathbb{S}_{j}^{\text{t}}\left(\mathbf{x}_{j,k}^{\text{t}}\right)$
of $\mathbb{X}_{0}$ and depends on $\mathbf{x}_{j,k}^{\text{t}}$.
When the target shape is rigid, \emph{e.g.}, for cars, $\mathbb{S}_{j}^{\text{t}}\left(\mathbf{x}_{j,k}^{\text{t}}\right)$
depends mainly on the target position and orientation. We assume that
\begin{equation}
\boldsymbol{x}_{j,k}^{\text{t}}\in\mathbb{S}_{j}^{\text{t}}\left(\mathbf{x}_{j,k}^{\text{t}}\right).\label{eq:xinShape}
\end{equation}
The target location $\boldsymbol{x}_{j,k}^{\text{t,g}}$, however,
does not necessarily belong to $\mathbb{S}_{j}^{\text{t}}\left(\mathbf{x}_{j,k}^{\text{t}}\right)$,
as in the case of cars, for example.

Usually, the type of targets to be localized is known. Consequently,
some information about the target dimensions is available. We assume
that for any target state $\mathbf{x}_{j,k}^{\text{t}}$ 
\begin{equation}
\mathbb{S}_{j}^{\text{t}}\left(\mathbf{x}_{j,k}^{\text{t}}\right)\subset\mathbb{C}^{\text{t}}\left(\boldsymbol{x}_{j,k}^{\text{t,g}}\right),\label{eq:ClassShape}
\end{equation}
where $\mathbb{C}^{\text{t}}\left(\boldsymbol{x}_{j,k}^{\text{t,g}}\right)$
is a vertical circular right cylinder of known height $h^{\text{t}}$
and radius $r^{\text{t}}$ with basis centered in $\boldsymbol{x}_{j,k}^{\text{t,g}}$.
The projection $\boldsymbol{p}_{\text{g}}\left(\mathbb{C}^{\text{t}}\left(\boldsymbol{x}_{j,k}^{\text{t,g}}\right)\right)$
of $\mathbb{C}^{\text{t}}\left(\boldsymbol{x}_{j,k}^{\text{t,g}}\right)$
on $\mathbb{X}_{\text{g}}$ is the disc $\mathbb{D}_{\text{g}}\left(\boldsymbol{x}_{j,k}^{\text{t,g}},r^{\text{t}}\right)$
of center $\boldsymbol{x}_{j,k}^{\text{t,g}}$ and radius $r^{\text{t}}$
included in $\mathbb{X}_{\text{g}}$.

We assume that targets cannot enter in any obstacle, \emph{i.e.},
\begin{equation}
\forall j\in\mathcal{N}^{\text{t}},\forall m\in\mathcal{N}^{\text{o}},\mathbb{S}_{j,k}^{\text{t}}\left(\mathbf{x}_{j,k}^{\text{t}}\right)\cap\mathbb{S}_{m}^{\text{o}}=\emptyset.\label{eq:SafeDistTargObs0}
\end{equation}

Consider the $r$-neighborhood of a set $\mathbb{S}\subset\mathbb{X}_{0}$
as
\begin{equation}
\mathbb{N}\left(\mathbb{S},r\right)=\left\{ \boldsymbol{x}\in\mathbb{X}_{0}\mid d\left(\boldsymbol{x},\mathbb{S}\right)\leqslant r\right\} \label{eq:rs_neighborhood}
\end{equation}
with $d\left(\boldsymbol{x},\mathbb{S}\right)=\min_{\boldsymbol{y}\in\mathbb{S}}\left\Vert \boldsymbol{x}-\boldsymbol{y}\right\Vert $
and the $r$-ground neighborhood of the projection on the ground of
a set $\mathbb{S}\subset\mathbb{X}_{0}$ defined as
\begin{equation}
\mathbb{N}_{\text{g}}\left(\mathbb{S},r\right)=\left\{ \boldsymbol{x}\in\mathbb{X}_{\text{g}}\mid d\left(\boldsymbol{x},\boldsymbol{p}_{\text{g}}\left(\mathbb{S}\right)\right)\leqslant r\right\} .\label{eq:GroundNeighborhood}
\end{equation}

\begin{figure}
\begin{centering}
\includegraphics[width=0.8\columnwidth]{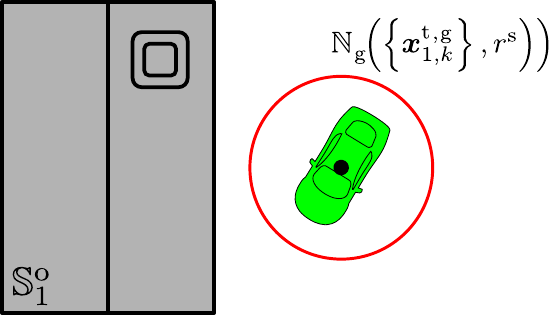}
\par\end{centering}
\caption{Obstacle (in grey), target (in green), and $r^{\text{s}}$-ground
neighborhood (in red) of its location. \label{fig:rsneighborhood}}
\end{figure}

We assume further that the target location $\boldsymbol{x}_{j,k}^{\text{t,g}}$
remains at a distance strictly larger than some known safety distance
$r^{\text{s}}\geqslant r^{\text{t}}$ of any obstacle $\mathbb{S}_{m}^{\text{o}}$,
\emph{i.e.}, that 
\begin{equation}
\forall j\in\mathcal{N}^{\text{t}},\forall m\in\mathcal{N}^{\text{o}},d\left(\boldsymbol{x}_{j,k}^{\text{t,g}},\mathbb{S}_{m}^{\text{o}}\right)>r^{\text{s}}.\label{eq:SafetyDistance-1}
\end{equation}
From~(\ref{eq:Obs_Zconvex}), one has $p_{\text{g}}\left(\mathbb{S}_{m}^{\text{o}}\right)\subset\mathbb{S}_{m}^{\text{o}}$.
Consequently, (\ref{eq:SafetyDistance-1}) implies that 
\begin{equation}
\forall j\in\mathcal{N}^{\text{t}},\forall m\in\mathcal{N}^{\text{o}},\mathbb{N}_{\text{g}}\left(\left\{ \boldsymbol{x}_{j,k}^{\text{t,g}}\right\} ,r^{\text{s}}\right)\cap\boldsymbol{p}_{\text{g}}\left(\mathbb{S}_{m}^{\text{o}}\right)=\emptyset,\label{eq:SafetyDistance}
\end{equation}
see Figure~\ref{fig:rsneighborhood}.

\subsection{UAV model\label{subsec:UAV-model}}

At time $t_{k}$, the state vector $\mathbf{x}_{i,k}^{\text{u}}$
of UAV~$i\in\mathcal{N}^{\text{u}}$ contains, among others, the
location of its center of gravity $\boldsymbol{x}_{i,k}^{\text{u}}\in\mathbb{R}^{2}\times\mathbb{R}^{+*}$.
The space occupied by UAV~$i$ is $\mathbb{S}^{\text{u}}\left(\mathbf{x}_{i,k}^{\text{u}}\right)$.
Its dynamic is modeled as
\begin{equation}
\mathbf{x}_{i,k+1}^{\text{u}}=\mathbf{f}^{\text{u}}\left(\mathbf{x}_{i,k}^{\text{u}},\mathbf{u}_{i,k}^{\text{u}}\right),\label{eq:UAV_Dyn}
\end{equation}
where $\mathbf{f}^{\text{u}}$ is known and the control input $\mathbf{u}_{i,k}^{\text{u}}$
is constrained within a bounded set $\mathbb{U}$. The state $\mathbf{x}_{i,k}^{\text{u}}$
is assumed to be perfectly known by UAV~$i$.

Each UAV is equipped by a camera and a CVS. In addition to the acquired
image $\mathbf{I}_{i,k}$ at time $t_{k}$, the CVS provides a depth-map
$\mathbf{D}_{i,k}$ obtained using, \emph{e.g.}, \cite{park_stereo_2012,madhuanand_self-supervised_2021,shimada_fast_2023},
an array of pixel labels $\mathbf{L}_{i,k}$ obtained using, \emph{e.g.},
\cite{grilli_review_2017,howard_searching_2019}, and a list $\mathcal{D}_{i,k}^{\text{t}}$
of identified targets. For each $j\in$$\mathcal{D}_{i,k}^{\text{t}}$,
a box $\left[\mathcal{Y}_{i,j,k}^{\text{t}}\right]$ in the image
$\mathbf{I}_{i,k}$ containing pixels of each identified target is
provided, \emph{e.g.}, by \cite{redmon_you_2016,minaeian_effective_2018,luo_yolod_2022,jiang_review_2022}.
These boxes are gathered in a list $\mathcal{B}_{i,k}^{\text{t}}=\left\{ \left[\mathcal{Y}_{i,j,k}^{\text{t}}\right]\right\} _{j\in\mathcal{D}_{i,k}}$.
We assume that $\mathbf{D}_{i,k}$ and $\mathbf{L}_{i,k}$ have the
same size as $\mathbf{I}_{i,k}$.

The communications within the fleet are modeled by the undirected
graph $\mathcal{G}_{k}=\left(\mathcal{N}^{\text{u}},\mathcal{E}_{k}\right)$,
with $\mathcal{N}^{\text{u}}$ the set of vertices and $\mathcal{E}_{k}=\mathcal{N}^{\text{u}}\times\mathcal{N}^{\text{u}}$
the set of edges of the graph. $\mathcal{E}_{k}$ describes the connectivity
at time $t_{k}$. For two UAVs $i\in\mathcal{N}^{\text{u}}$ and $i'\in\mathcal{N}^{\text{u}}$,
if $\left(i,i'\right)\in\mathcal{E}_{k}$, then both UAVs are able
to exchange information without delay and error. The set $\mathcal{N}_{i,k}=\left\{ i'\in\mathcal{N}^{\text{u}}\mid\left(i,i'\right)\in\mathcal{E}_{k}\right\} $
contains all UAV indexes with which UAV~$i$ is able to communicate
at time $t_{k}$. We consider that $i\in\mathcal{N}_{i,k}$.

\subsection{Problem formulation\label{subsec:Problem-formulation}}

The information available to UAV~$i$ before time $t_{k}$ is gathered
in the set $\mathcal{I}_{i,k-1}$. $\mathcal{I}_{i,k-1}$ contains,
among others, a list $\mathcal{L}_{i,k-1}^{\text{t}}$ of indexes
of targets already identified. For each $j\in\mathcal{L}_{i,k-1}^{\text{t}}$,
we assume that UAV~$i$ has access to a set estimate $\mathbb{X}_{i,j,k-1}^{\text{t}}\subset\mathbb{X}_{\text{g}}$
of all possible target locations $\boldsymbol{x}_{j,k-1}^{\text{t,g}}$
which are consistent with $\mathcal{I}_{i,k-1}$. It has also access
to a set $\overline{\mathbb{X}}_{i,k-1}^{\text{t}}$ containing all
locations where targets still to be detected may be present.

Using the information collected by the camera and processed by the
CVS at time $t_{k}$, the information available at time $t_{k}$ to
UAV~$i$ is then $\mathcal{I}_{i,k|k}=\mathcal{I}_{i,k-1}\cup\left\{ \mathbf{I}_{i,k},\mathbf{D}_{i,k},\mathbf{L}_{i,k},\mathcal{D}_{i,k}^{\text{t}},\mathcal{B}_{i,k}^{\text{t}}\right\} .$
With this information UAV~$i$ evaluates an updated list of identified
targets $\mathcal{L}_{i,k|k}^{\text{t}}=\mathcal{L}_{i,k-1}^{\text{t}}\cup\mathcal{D}_{i,k}^{\text{t}}$.
UAV~$i$ has then, for each target $j\in\mathcal{L}_{i,k|k}^{\text{t}}$,
to characterize the set $\mathbb{X}_{i,j,k|k}^{\text{t}}\subset\mathbb{X}_{\text{g}}$
of all possible target locations $\boldsymbol{x}_{j,k}^{\text{t,g}}$
which are consistent with $\mathcal{I}_{i,k|k}$. UAV~$i$ has also
to update the estimate $\overline{\mathbb{X}}_{i,k-1}^{\text{t}}$
to get $\overline{\mathbb{X}}_{i,k|k}^{\text{t}}$.

Then, UAV~$i$ broadcasts some updated information to its neighbors
$\mathcal{N}_{i,k}$ and receives information from them. The type
of information exchanged is detailed in Section~\ref{subsec:Correc-comm}.
After communication, the total information UAV~$i$ has now access
to is denoted $\mathcal{I}_{i,k}$. UAV~$i$ can then further update
its list of identified targets to get $\mathcal{L}_{i,k}^{\text{t}}$,
the set estimates $\mathbb{X}_{i,j,k}^{\text{t}}$ for each $j\in\mathcal{L}_{i,k}^{\text{t}}$
gathered in the set $\mathcal{X}_{i,k}^{\text{t}}=\left\{ \mathbb{X}_{i,j,k}^{\text{t}}\right\} _{j\in\mathcal{L}_{i,k}^{\text{t}}}$,
and evaluate an updated version of $\overline{\mathbb{X}}_{i,k|k}^{\text{t}}$
denoted $\overline{\mathbb{X}}_{i,k}^{\text{t}}$.

The target localization uncertainty is defined as
\begin{equation}
\Phi\left(\mathcal{X}_{i,k}^{\text{t}},\overline{\mathbb{X}}_{i,k}^{\text{t}}\right)=\phi\left(\overline{\mathbb{X}}_{i,k}^{\text{t}}\cup{\textstyle \bigcup}_{j\in\mathcal{L}_{i,k}^{\text{t}}}\mathbb{X}_{i,j,k}^{\text{t}}\right),\label{eq:TargLocUncertaintyEstimation}
\end{equation}
where $\phi\left(\mathbb{X}\right)$ is the area of the set $\mathbb{X}\subset\mathbb{R}^{2}$.
$\Phi\left(\mathcal{X}_{i,k}^{\text{t}},\overline{\mathbb{X}}_{i,k}^{\text{t}}\right)$
account for already identified and still to be identified targets.

Our aim in what follows is (\emph{i}) to show the way set estimates
are obtained from CVS information, (\emph{ii}) to determine the type
of information to be exchanged between UAVs, (\emph{iii}) to present
the way set estimates are updated using $\mathcal{I}_{i,k}^{\mathcal{N}}$,
and (\emph{iv}) to design the trajectory of each UAV so as to minimize
(\ref{eq:TargLocUncertaintyEstimation}).

\section{Hypotheses on the CVS\label{sec:HypCVS}}

In this section, some hypotheses are introduced to exploit the information
provided by the CVS using set-membership estimation techniques. Section~\ref{subsec:Camera-model}
presents a geometrical model for the camera embedded in UAVs. Then,
Section~\ref{subsec:Exploiting-information} introduces hypotheses
related to CVS information. Section~\ref{subsec:Targ_ObservationAssump}
presents an assumption on the light rays illuminating the CCD array
when the location of a target is in the FoV of an UAV. In the remainder
of this section, the time index $k$ is omitted to lighten the notations.

\subsection{Camera model and field of view\label{subsec:Camera-model}}

\begin{figure}
\begin{centering}
\includegraphics[width=1\columnwidth]{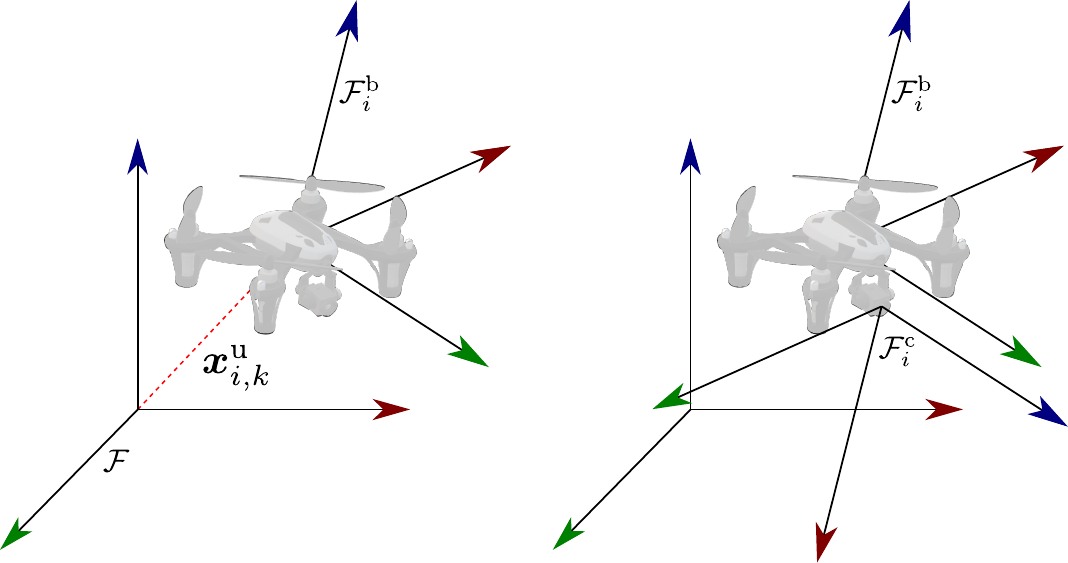}
\par\end{centering}
\caption{\label{fig:UAV_BodyFrame}Left: Reference frame $\mathcal{F}$ and
UAV body frame $\mathcal{F}_{i}^{\text{b}}$; Right: Camera frame
$\mathcal{F}_{i}^{\text{c}}$ and body frame $\mathcal{F}_{i}^{\text{b}}$
of UAV~$i$, when $\theta=0$}
\end{figure}

A body frame $\mathcal{F}_{i}^{\text{b}}$, illustrated in Figure~\ref{fig:UAV_BodyFrame},
with origin $\boldsymbol{x}_{i,k}^{\text{u}}$ is attached to UAV~$i$.
The rotation matrix from $\mathcal{F}$ to $\mathcal{F}_{i}^{\text{b}}$
is denoted $\mathbf{M}_{\mathcal{F}}^{\mathcal{F}_{i}^{\text{b}}}$.
$\mathbf{M}_{\mathcal{F}}^{\mathcal{F}_{i}^{\text{b}}}$ depends on
$\mathbf{x}_{i,k}^{\text{u}}$. The coordinates of some vector $\boldsymbol{x}\in\mathbb{R}^{3}$
in $\mathcal{F}$, when expressed in $\mathcal{F}_{i}^{\text{b}}$,
are
\begin{equation}
\mathbf{T}_{\mathcal{F}}^{\mathcal{F}_{i}^{\text{b}}}\left(\boldsymbol{x}\right)=\mathbf{M}_{\mathcal{F}}^{\mathcal{F}_{i}^{\text{b}}}\left(\boldsymbol{x}-\boldsymbol{x}_{i,k}^{\text{u}}\right).\label{eq:Transform_F_Fb}
\end{equation}

Each UAV~$i$ of the fleet embeds the same type of CCD camera which
provides an image $\mathbf{I}_{i}$ of $N_{\text{r}}$ rows and $N_{\text{c}}$
columns. The camera frame $\mathcal{F}_{i}^{\text{c}}$ of UAV~$i$
has its optical center $\boldsymbol{x}_{i}^{\text{c},\mathcal{F}_{i}^{\text{b}}}$
as origin and is oriented such that the positive $z$-axis represents
the optical axis of the camera. The $x$-axis is parallel to the pixel
rows in the CCD array and the $y$-axis is parallel to the columns.
The rotation matrix from $\mathcal{F}_{i}^{\text{b}}$ to $\mathcal{F}_{i}^{\text{c}}$
is
\begin{equation}
\mathbf{M}_{\mathcal{F}_{i}^{\text{b}}}^{\mathcal{F}_{i}^{\text{c}}}=\begin{pmatrix}0 & -1 & 0\\
0 & 0 & -1\\
1 & 0 & 0
\end{pmatrix}\begin{pmatrix}\cos\theta & 0 & -\sin\theta\\
0 & 1 & 0\\
\sin\theta & 0 & \cos\theta
\end{pmatrix},
\end{equation}
where $\theta$ is the fixed camera angle between the $x$-axis of
$\mathcal{F}_{i}^{\text{b}}$ and the $z$-axis of $\mathcal{F}_{i}^{\text{c}}$.
Figure~\ref{fig:UAV_BodyFrame} illustrates these two frames.

The dimensions $H_{\text{r}}\times H_{\text{c}}$ of the CCD array
of the camera and its focal length $f$ are known. A pinhole model
without distortion \cite{faugeras_three-dimensional_1993} is considered
for the camera. Therefore, the matrix of intrinsic parameters \cite{faugeras_three-dimensional_1993}
used to project a point of $\mathcal{F}_{i}^{\text{c}}$ onto the
CCD array is
\begin{equation}
\mathbf{K}=\begin{pmatrix}-f_{\text{c}} & 0 & N_{\text{c}}/2\\
0 & -f_{\text{r}} & N_{\text{r}}/2
\end{pmatrix}\label{eq:ProjectionMatrix_Camera}
\end{equation}
where $f_{\text{c}}=fN_{\text{c}}/H_{\text{c}}$ and $f_{\text{r}}=fN_{\text{r}}/H_{\text{r}}$are
the focal lengths expressed in pixels. Consequently, the 2D coordinates
on the CCD array of the projection of some point $\boldsymbol{x}^{\mathcal{F}_{i}^{\text{c}}}\in\mathbb{R}^{3}$
are
\begin{align}
\boldsymbol{p}_{\mathcal{F}_{i}^{\text{c}}}\left(\boldsymbol{x}^{\mathcal{F}_{i}^{\text{c}}}\right) & =\mathbf{K}\boldsymbol{x}^{\mathcal{F}_{i}^{\text{c}}}/x_{3}^{\mathcal{F}_{i}^{\text{c}}}.\label{eq:Pinhole_model}
\end{align}

According to the considered pinhole model, each light ray passing
through the optical center of the camera and illuminating the CCD
array at $\left(x,y\right)\in\left[0,N_{\text{c}}\right]\times\left[0,N_{r}\right]$
can be modeled by a half-line of direction
\begin{equation}
\boldsymbol{v}^{\mathcal{F}_{i}^{\text{c}}}\left(x,y\right)=\frac{1}{\nu\left(x,y\right)}\left(\begin{array}{c}
\left(N_{\text{c}}/2-x\right)/f_{\text{c}}\\
\left(N_{\text{r}}/2-y\right)/f_{\text{r}}\\
1
\end{array}\right),\label{eq:LightRay_UnitVector}
\end{equation}
with {\small{}$\nu\left(x,y\right)={\textstyle \sqrt{\left(\left(N_{\text{c}}/2-x\right)/f_{\text{c}}\right)^{2}+\left(\left(N_{\text{r}}/2-y\right)/f_{\text{r}}\right)^{2}+1}}$}.
The set
\begin{align}
\mathcal{V}_{i}\left(n_{\text{r}},n_{\text{c}}\right)= & \left\{ \mathbf{M}_{\mathcal{F}_{i}^{\text{c}}}^{\mathcal{F}}\boldsymbol{v}^{\mathcal{F}_{i}^{\text{c}}}\left(x,y\right)\mid\right.\nonumber \\
 & \left.x\in\left[n_{\text{c}}-1,n_{\text{c}}\right],y\in\left[n_{\text{r}}-1,n_{\text{r}}\right]\right\} ,\label{eq:Set_LoS_Vi}
\end{align}
with $\mathbf{M}_{\mathcal{F}_{i}^{\text{c}}}^{\mathcal{F}}=\mathbf{M}_{\mathcal{F}_{i}^{\text{b}}}^{\mathcal{F}}\mathbf{M}_{\mathcal{F}_{i}^{\text{c}}}^{\mathcal{F}_{i}^{\text{b}}}$,
contains the directions, expressed in $\mathcal{F}$, of all light
rays contributing to the illumination of the pixel $\left(n_{\text{r}},n_{\text{c}}\right)$.

The FoV $\mathbb{F}\left(\mathbf{x}_{i}^{\text{u}}\right)$ of the
CCD camera of UAV~$i$ (dubbed as FoV of UAV~$i$ in what follows)
represents the set of points in $\mathbb{X}_{0}$ that are potentially
observed (ignoring obstacles and targets). $\mathbb{F}\left(\mathbf{x}_{i}^{\text{u}}\right)$
is a half-cone with the camera optical center $\boldsymbol{x}_{i}^{\text{c}}$
as its apex and with four unit vectors $\boldsymbol{v}_{\ell}^{\mathcal{F}_{i}^{\text{c}}}$,
$\ell=1,\dots,4$, describing its edges, \emph{i.e.},
\begin{equation}
\mathbb{F}\left(\mathbf{x}_{i}^{\text{u}}\right)=\left\{ \boldsymbol{x}_{i}^{\text{c}}+{\textstyle \sum_{\ell=1}^{4}a_{\ell}\mathbf{M}_{\mathcal{F}_{i}^{\text{c}}}^{\mathcal{F}}\boldsymbol{v}_{\ell}^{\mathcal{F}_{i}^{\text{c}}}\ |\ a_{\ell}\in\mathbb{R}^{+}}\right\} .\label{eq:FoV_halfCone}
\end{equation}
These four unit vectors can be deduced from (\ref{eq:LightRay_UnitVector})
by taking $\left(x,y\right)$ at the four corners of the CCD array.
In what follows, we assume that any information related to some $\boldsymbol{x}\in\mathbb{F}\left(\mathbf{x}_{i}^{\text{u}}\right)$
at a distance from $\boldsymbol{x}_{i}^{\text{c}}$ larger than $d_{\max}$
cannot be considered as reliable. One introduces then the subset
\begin{equation}
\underline{\mathbb{F}}\left(\mathbf{x}_{i}^{\text{u}}\right)=\mathbb{F}\left(\mathbf{x}_{i}^{\text{u}}\right)\cap\mathbb{B}\left(\boldsymbol{x}_{i}^{\text{c}},d_{\max}\right),\label{eq:ReliablePointInFOV}
\end{equation}
 of the FoV where $\mathbb{B}\left(\boldsymbol{x}_{i}^{\text{c}},d_{\max}\right)$
is the ball of $\mathbb{R}^{3}$ of center $\boldsymbol{x}_{i}^{\text{c}}$
and radius $d_{\max}$. $\underline{\mathbb{F}}\left(\mathbf{x}_{i}^{\text{u}}\right)$
contains all points of $\mathbb{F}\left(\mathbf{x}_{i}^{\text{u}}\right)$
at a distance from $\boldsymbol{x}_{i}^{\text{c}}$ less than $d_{\max}$.

The coordinates of a point $\boldsymbol{x}\in\mathbb{F}\left(\mathbf{x}_{i}^{\text{u}}\right)$,
when expressed in $\mathcal{F}_{i}^{\text{b}}$ are $\boldsymbol{x}^{\mathcal{F}_{i}^{\text{b}}}=\mathbf{T}_{\mathcal{F}}^{\mathcal{F}_{i}^{\text{b}}}\left(\boldsymbol{x}\right)\in\mathbb{R}^{3}$.
To express them in $\mathcal{F}_{i}^{\text{c}}$, one evaluates
\begin{equation}
\mathbf{T}_{\mathcal{F}_{i}^{\text{b}}}^{\mathcal{F}_{i}^{\text{c}}}\left(\boldsymbol{x}^{\mathcal{F}_{i}^{\text{b}}}\right)=\mathbf{M}_{\mathcal{F}_{i}^{\text{b}}}^{\mathcal{F}_{i}^{\text{c}}}\left(\boldsymbol{x}^{\mathcal{F}_{i}^{\text{b}}}-\boldsymbol{x}_{i}^{\text{c},\mathcal{F}_{i}^{\text{b}}}\right).\label{eq:Transform_Fb_Fc}
\end{equation}
Composing (\ref{eq:Transform_F_Fb}) and (\ref{eq:Transform_Fb_Fc}),
one gets the transform 
\begin{equation}
\mathbf{T}_{\mathcal{F}}^{\mathcal{F}_{i}^{\text{c}}}=\mathbf{T}_{\mathcal{F}_{i}^{\text{b}}}^{\mathcal{F}_{i}^{\text{c}}}\circ\mathbf{T}_{\mathcal{F}}^{\mathcal{F}_{i}^{\text{b}}}\label{eq:CameraModel_ChgtFrame}
\end{equation}
to express the coordinates of $\boldsymbol{x}\in\mathbb{F}\left(\mathbf{x}_{i}^{\text{u}}\right)$
in $\mathcal{F}_{i}^{\text{c}}$. Combining (\ref{eq:Pinhole_model})
and (\ref{eq:CameraModel_ChgtFrame}), the coordinates of the image
on the CCD array of a point $\boldsymbol{x}\in\mathbb{F}\left(\mathbf{x}_{i}^{\text{u}}\right)$
are 
\begin{equation}
\boldsymbol{p}_{\text{c}}\left(\mathbf{x}_{i}^{\text{u}},\boldsymbol{x}\right)=\boldsymbol{p}_{\mathcal{F}_{i}^{\text{c}}}\left(\mathbf{T}_{\mathcal{F}}^{\mathcal{F}_{i}^{\text{c}}}\left(\boldsymbol{x}\right)\right).
\end{equation}

The notation $\boldsymbol{p}_{\text{c}}\left(\mathbf{x}_{i}^{\text{u}},\boldsymbol{x}\right)\in\left(n_{\text{r}},n_{\text{c}}\right)$,
with $\boldsymbol{x}\in\mathbb{F}\left(\mathbf{x}_{i}^{\text{u}}\right)$,
indicates that the projection of $\boldsymbol{x}$ on the CCD array
belongs to the pixel with coordinates $\left(n_{\text{r}},n_{\text{c}}\right)\in\mathcal{N}^{\text{I}}$,
with $\mathcal{N}^{\text{I}}=\left\{ 1\dots N_{\text{r}}\right\} \times\left\{ 1\dots N_{\text{c}}\right\} $.
Consequently, one has
\begin{equation}
\boldsymbol{x}\in\mathbb{F}\left(\mathbf{x}_{i}^{\text{u}}\right)\Rightarrow\exists\left(n_{\text{r}},n_{\text{c}}\right)\in\mathcal{N}^{\text{I}},\boldsymbol{p}_{\text{c}}\left(\mathbf{x}_{i}^{\text{u}},\boldsymbol{x}\right)\in\left(n_{\text{r}},n_{\text{c}}\right).\label{eq:PixelExistence4PointsInFoV}
\end{equation}

\begin{figure}
\begin{centering}
\includegraphics[width=0.8\columnwidth]{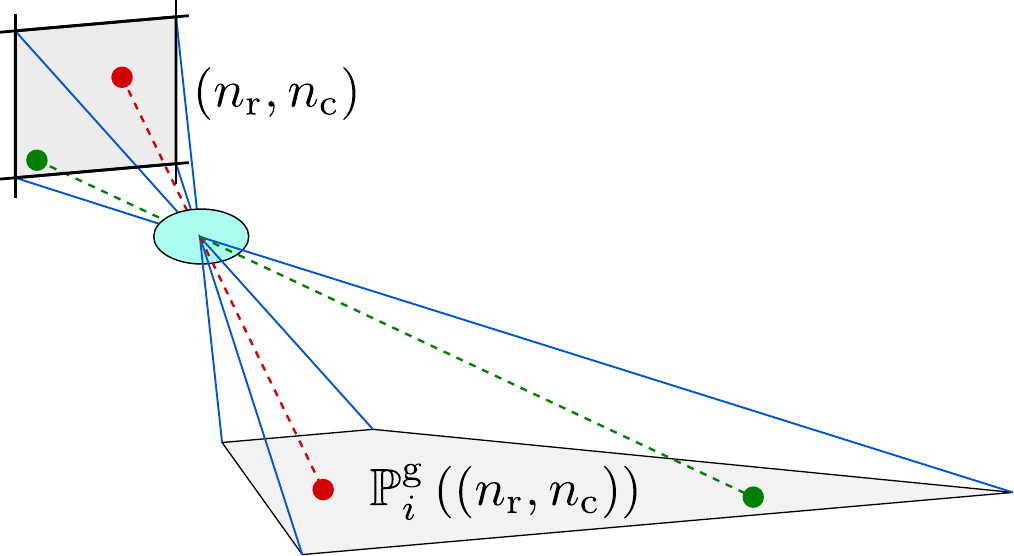}
\par\end{centering}
\caption{\label{fig:Pinhole-model}Pinhole model of the camera and several
light rays contributing to the illumination of the pixel $\left(n_{\text{r}},n_{\text{c}}\right)$;
The $4$ blue lines are the light rays illuminating the corners of
$\left(n_{\text{r}},n_{\text{c}}\right)$; The quadrangle $\mathbb{P}_{i}^{\text{g}}\left(\left(n_{\text{r}},n_{\text{c}}\right)\right)$
represents all points of the ground which may contribute to the illumination
of pixel $\left(n_{\text{r}},n_{\text{c}}\right)$, see Section~\ref{sec:ExploitingCVS}.}
\end{figure}

\subsection{Exploiting information provided by the CVS\label{subsec:Exploiting-information}}

This section presents the assumptions considered to exploit the information
provided by the CVS embedded in each UAV. Section~\ref{subsec:Depth-map-model}
introduces an assumption on the information in $\mathbf{D}_{i}$ used
to get bounded-error measurements of the distance from the camera
to parts of the environment. Section~\ref{subsec:Classifier} presents
assumptions on the pixel labels $\mathbf{L}_{i}$. Section~\ref{subsec:Targ_Detection}
introduces assumptions related to bounding boxes $\left[\mathcal{Y}_{i,j}^{\text{t}}\right],$$j\in\mathcal{D}_{i},$
and labeled pixels for targets identified by the CVS.

\subsubsection{From a depth map to bounded-error range measurements\label{subsec:Depth-map-model}}

The distance between the optical center of the camera $\boldsymbol{x}_{i}^{\text{c}}$
and the environment (including obstacles and targets) along $\boldsymbol{v}\in\mathcal{V}_{i}\left(n_{\text{r}},n_{\text{c}}\right)$,
for any $\left(n_{\text{r}},n_{\text{c}}\right)\in\mathcal{N}^{\text{I}}$,
is
\begin{align}
\rho\left(\boldsymbol{x}_{i}^{\text{c}},\boldsymbol{v}\right)= & \min\left\{ d_{\boldsymbol{v}}\left(\boldsymbol{x}_{i}^{\text{c}},\mathbb{X}_{\text{g}}\right),d_{\boldsymbol{v}}\left(\boldsymbol{x}_{i}^{\text{c}},{\textstyle \bigcup}_{m\in\mathcal{N}^{\text{o}}}\mathbb{S}_{m}^{\text{o}}\right),\right.\label{eq:range_model}\\
 & \left.d_{\boldsymbol{v}}\left(\boldsymbol{x}_{i}^{\text{c}},{\textstyle \bigcup}_{j\in\mathcal{N}^{\text{t}}}\mathbb{S}_{j}^{\text{t}}\left(\mathbf{x}_{j}^{\text{t}}\right)\right),d_{\boldsymbol{v}}\left(\boldsymbol{x}_{i}^{\text{c}},{\textstyle \bigcup}_{\ell\in\mathcal{N}^{\text{u}}}\mathbb{S}^{\text{u}}\left(\mathbf{x}_{\ell}^{\text{u}}\right)\right)\right\} ,\nonumber 
\end{align}
where $d_{\boldsymbol{v}}\left(\boldsymbol{x},\mathbb{S}\right)$
is the distance from a point $\boldsymbol{x}\in\mathbb{X}_{0}$ to
the intersection of the set $\mathbb{S}$ along the half-line of origin
$\boldsymbol{x}$ and direction $\boldsymbol{v}$. Then
\begin{equation}
\rho\left(\boldsymbol{x}_{i}^{\text{c}},\mathcal{V}_{i}\left(n_{\text{r}},n_{\text{c}}\right)\right)=\left\{ \rho\left(\boldsymbol{x}_{i}^{\text{c}},\boldsymbol{v}\right)\mid\boldsymbol{v}\in\mathcal{V}_{i}\left(n_{\text{r}},n_{\text{c}}\right)\right\} ,\label{eq:DepthMap_DepthMap}
\end{equation}
is the set of all distances between $\boldsymbol{x}_{i}^{\text{c}}$
and the environment along any direction $\boldsymbol{v}\in\mathcal{V}_{i}\left(n_{\text{r}},n_{\text{c}}\right)$.

\begin{figure}
\begin{centering}
\includegraphics[width=0.8\columnwidth]{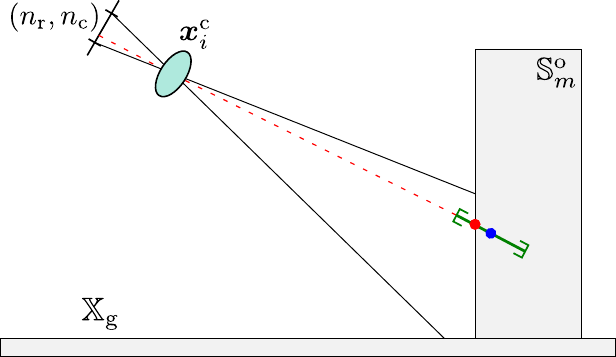}
\par\end{centering}
\caption{Depth map information for one pixel: $\mathbf{D}_{i}^{0}\left(n_{\text{r}},n_{\text{c}}\right)$
is the distance between $\boldsymbol{x}_{i}^{\text{c}}$ and the red
dot, while $\mathbf{D}_{i}\left(n_{\text{r}},n_{\text{c}}\right)$
represents the measured distance between $\boldsymbol{x}_{i}^{\text{c}}$
and the blue dot; The green interval is $\left[\mathbf{D}_{i}\right]\left(n_{\text{r}},n_{\text{c}}\right)$.\label{fig:Depth-map-information}}
\end{figure}

We assume that each element $\mathbf{D}_{i}\left(n_{\text{r}},n_{\text{c}}\right)$
of the depth map is a noisy version 
\begin{equation}
\mathbf{D}_{i}\left(n_{\text{r}},n_{\text{c}}\right)=\mathbf{D}_{i}^{0}\left(n_{\text{r}},n_{\text{c}}\right)\left(1+w\right)\label{eq:D=00003DD0(1+w)}
\end{equation}
of the distance $\mathbf{D}_{i}^{0}\left(n_{\text{r}},n_{\text{c}}\right)=\rho\left(\mathbf{x}_{i}^{\text{u}},\boldsymbol{v}\right)$
between $\boldsymbol{x}_{i}^{\text{c}}$ and the environment along
some unknown direction $\boldsymbol{v}\in\mathcal{V}_{i}\left(n_{\text{r}},n_{\text{c}}\right)$.
In (\ref{eq:D=00003DD0(1+w)}), the noise $w$ is assumed to belong
to the known interval $\left[\underline{w},\overline{w}\right]$.
Consequently, the interval
\begin{equation}
\left[\mathbf{D}_{i}\right]\left(n_{\text{r}},n_{\text{c}}\right)=\left[1/\left(1+\overline{w}\right),1/\left(1+\underline{w}\right)\right]\mathbf{D}_{i}\left(n_{\text{r}},n_{\text{c}}\right).\label{eq:DepthMap_Interval}
\end{equation}
contains $\mathbf{D}_{i}^{0}\left(n_{\text{r}},n_{\text{c}}\right)$,
see Figure~\ref{fig:Depth-map-information}.

\subsubsection{Pixel classification\label{subsec:Classifier}}

Each element $\left(n_{\text{r}},n_{\text{c}}\right)$ of the array
of pixel labels $\mathbf{L}_{i}$ is assumed to belong to one of the
following classes: Ground, Target, Obstacle, and Unknown/Not Labeled.
The latter class corresponds to pixels that cannot be classified in
one of the three other classes due to a lack of confidence. According
to (\ref{eq:ReliablePointInFOV}), all information coming from a point
that may not be in $\underline{\mathbb{F}}\left(\mathbf{x}_{i}^{\text{u}}\right)$
is considered as unreliable. Therefore, using $\mathbf{D}_{i}$, one
introduces the list of pixel coordinates 
\begin{equation}
\mathcal{Y}_{i}=\left\{ \left(n_{\text{r}},n_{\text{c}}\right)\in\mathcal{N}^{\text{I}}\mid\frac{1}{1+\underline{w}}\mathbf{D}_{i}\left(n_{\text{r}},n_{\text{c}}\right)\leqslant d_{\max}\right\} \label{eq:reliablePixels}
\end{equation}
for which reliable information is assumed available, especially about
their labeling. All pixels with indexes in $\mathcal{Y}_{i}$ are
assumed to be correctly classified (possibly as Unknown, when there
is an ambiguity).

\begin{figure}
\begin{centering}
\includegraphics[width=0.8\columnwidth]{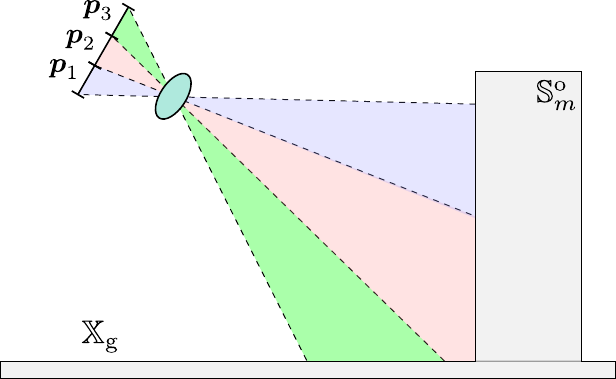}
\par\end{centering}
\caption{As pixel $\boldsymbol{p}_{1}\in\mathcal{Y}_{i}^{\text{o}}$, all light
rays illuminating $\boldsymbol{p}_{1}$ stem from an obstacle; as
pixel $\boldsymbol{p}_{3}\in\mathcal{Y}_{i}^{\text{g}}$, all light
rays illuminating $\boldsymbol{p}_{3}$ stem from $\mathbb{X}_{\text{g}}$;
As $\boldsymbol{p}_{2}\in\mathcal{Y}_{i}^{\text{n}}$, nothing can
be concluded.\label{fig:Pixel_classif}}
\end{figure}

Four subsets of $\mathcal{Y}_{i}$ are then deduced from $\mathbf{L}_{i}$,
namely $\mathcal{Y}_{i}^{\text{g}}$, $\mathcal{Y}_{i}^{\text{t}}$,
$\mathcal{Y}_{i}^{\text{o}}$, and $\mathcal{Y}_{i}^{\text{n}}$ gathering
coordinates of pixels respectively labeled as Ground, Target, Obstacle,
and Unknown/Not Labeled. We assume that pixels corresponding to other
UAVs are labeled as Unknown. Moreover, we assume that if a pixel $\left(n_{\text{r}},n_{\text{c}}\right)\in\mathcal{Y}_{i}^{\text{g}}$,
then all light rays illuminating $\left(n_{\text{r}},n_{\text{c}}\right)$
stem from the ground $\mathbb{X}_{\text{g}}$, \emph{i.e.}, 
\begin{equation}
\forall\boldsymbol{v}\in\mathcal{V}_{i}\left(n_{\text{r}},n_{\text{c}}\right),\rho\left(\boldsymbol{x}_{i}^{\text{c}},\boldsymbol{v}\right)=d_{\boldsymbol{v}}\left(\boldsymbol{x}_{i}^{\text{c}},\mathbb{X}_{\text{g}}\right).\label{eq:Classif_GroundPix}
\end{equation}
Similarly, if a pixel $\left(n_{\text{r}},n_{\text{c}}\right)\in\mathcal{Y}_{i}^{\text{t}}$,
then there exists a target~$j\in\mathcal{N}^{\text{t}}$ such that
all light rays illuminating $\left(n_{\text{r}},n_{\text{c}}\right)$
stem from $\mathbb{S}_{j}^{\text{t}}\left(\mathbf{x}_{j,k}^{\text{t}}\right)$,
\emph{i.e.}, 
\begin{equation}
\forall\boldsymbol{v}\in\mathcal{V}_{i}\left(n_{\text{r}},n_{\text{c}}\right),\exists j\in\mathcal{N}^{\text{t}},\rho\left(\boldsymbol{x}_{i}^{\text{c}},\boldsymbol{v}\right)=d_{\boldsymbol{v}}\left(\boldsymbol{x}_{i}^{\text{c}},\mathbb{S}_{j}^{\text{t}}\left(\mathbf{x}_{j,k}^{\text{t}}\right)\right).\label{eq:Classif_TargPix}
\end{equation}
Finally, if a pixel $\left(n_{\text{r}},n_{\text{c}}\right)\in\mathcal{Y}_{i}^{\text{o}}$,
then for all light rays illuminating $\left(n_{\text{r}},n_{\text{c}}\right)$,
there exists an obstacle~$m\in\mathcal{N}^{\text{o}}$ such that
the light ray stem from $\mathbb{S}_{m}^{\text{o}}$, \emph{i.e.},
\begin{equation}
\forall\boldsymbol{v}\in\mathcal{V}_{i}\left(n_{\text{r}},n_{\text{c}}\right),\exists m\in\mathcal{N}^{\text{o}},\rho\left(\boldsymbol{x}_{i}^{\text{c}},\boldsymbol{v}\right)=d_{\boldsymbol{v}}\left(\boldsymbol{x}_{i}^{\text{c}},\mathbb{S}_{m}^{\text{o}}\right).\label{eq:Classif_ObsPix-1}
\end{equation}
See Figure~\ref{fig:Pixel_classif}.

\subsubsection{Bounding boxes for identified targets\label{subsec:Targ_Detection}}

When $\mathcal{Y}_{i}^{\text{t}}$ is not empty, at least one target
located within $\mathbb{F}\left(\mathbf{x}_{i}^{\text{u}}\right)$
has been detected. In such case, the CVS may also provide a list $\mathcal{D}_{i}^{\text{t}}\subset\mathcal{N}^{\text{t}}$
of identified targets and an axis-aligned box $\left[\mathcal{Y}_{i,j}^{\text{t}}\right]$,
for each $j\in\mathcal{D}_{i}^{\text{t}}$. $\mathcal{D}_{i}^{\text{t}}$
may be empty even if $\mathcal{Y}_{i}^{\text{t}}$ is not empty, when
$\mathcal{Y}_{i}^{\text{t}}$ does not contain enough information
to identify a target.

Consider the set $\mathcal{Y}_{i,j}^{\text{t}}\subset\mathcal{Y}_{i}^{\text{t}}$
containing all pixels of $\mathcal{Y}_{i}^{\text{t}}$ associated
to target~$j$ only. If $j\in\mathcal{D}_{i}^{\text{t}}$, we assume
that $\mathcal{Y}_{i,j}^{\text{t}}$ is not empty, \emph{i.e.},
\begin{equation}
j\in\mathcal{D}_{i}^{\text{t}}\Rightarrow\mathcal{Y}_{i,j}^{\text{t}}\neq\emptyset.\label{eq:TargDetection_Yij}
\end{equation}
Moreover, the CVS is assumed to be tuned in such a way that $\left[\mathcal{Y}_{i,j}^{\text{t}}\right]$
contains $\mathcal{Y}_{i,j}^{\text{t}}$, \emph{i.e.},
\begin{equation}
j\in\mathcal{D}_{i}^{\text{t}}\Rightarrow\mathcal{Y}_{i,j}^{\text{t}}\subset\left[\mathcal{Y}_{i,j}^{\text{t}}\right].\label{eq:BoxIntersectsYt}
\end{equation}
We assume further that the classifier is unable to provide $\mathcal{Y}_{i,j}^{\text{t}}$.
Therefore, the target location estimator will have to exploit $\left[\mathcal{Y}_{i,j}^{\text{t}}\right]$
and $\mathcal{Y}_{i}^{\text{t}}$ only.

\subsection{Assumption on observed targets\label{subsec:Targ_ObservationAssump}}

\begin{figure}
\begin{centering}
\includegraphics[width=0.8\columnwidth]{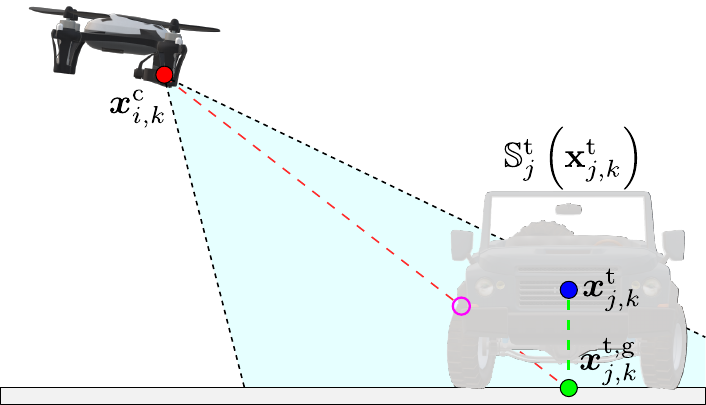}
\par\end{centering}
\caption{\label{fig:Assumption_TargObs}Illustration of (\ref{eq:RayonIntersection}):
The blue and green dots represent respectively $\boldsymbol{x}_{j,k}^{\text{t}}$
and $\boldsymbol{x}_{j,k}^{\text{t,g}}$; The dashed red line is the
line-of-sight $\left[\boldsymbol{x}_{i,k}^{\text{c}},\boldsymbol{x}_{j,k}^{\text{t,g}}\right[$;
The purple dot represents the part of $\mathbb{S}_{j}^{\text{t}}\left(\mathbf{x}_{j}^{\text{t}}\right)$
seen by the UAV along $\left[\boldsymbol{x}_{i,k}^{\text{c}},\boldsymbol{x}_{j,k}^{\text{t,g}}\right[$.}
\end{figure}

Consider some target $j$ such that $\boldsymbol{x}_{j}^{\text{t,g}}\in\underline{\mathbb{F}}\left(\mathbf{x}_{i}^{\text{u}}\right)$.
We assume that the half-open segment $\left[\boldsymbol{x}_{i}^{\text{c}},\boldsymbol{x}_{j}^{\text{t,g}}\right[$
intersects $\mathbb{S}_{j}^{\text{t}}\left(\mathbf{x}_{j}^{\text{t}}\right)$,
\emph{i.e.}, 
\begin{equation}
\boldsymbol{x}_{j}^{\text{t,g}}\in\underline{\mathbb{F}}\left(\mathbf{x}_{i}^{\text{u}}\right)\implies\left[\boldsymbol{x}_{i}^{\text{c}},\boldsymbol{x}_{j}^{\text{t,g}}\right[\cap\mathbb{S}_{j}^{\text{t}}\left(\mathbf{x}_{j}^{\text{t}}\right)\neq\emptyset.\label{eq:RayonIntersection}
\end{equation}
Thus, if $\boldsymbol{x}_{j}^{\text{t,g}}$ belongs to the FoV, then
some points on $\mathbb{S}_{j}^{\text{t}}\left(\mathbf{x}_{j}^{\text{t}}\right)$
will reflect a light ray that will illuminate the CCD array of the
camera (in absence of obstacles), see Figure~\ref{fig:Assumption_TargObs}.
Consequently, if $\boldsymbol{x}_{j}^{\text{t,g}}\in\underline{\mathbb{F}}\left(\mathbf{x}_{i}^{\text{u}}\right)$,
then there exists a pixel $\left(n_{\text{r}},n_{\text{c}}\right)$
such that $\boldsymbol{p}_{\text{c}}\left(\mathbf{x}_{i}^{\text{u}},\boldsymbol{x}_{j}^{\text{t,g}}\right)\in\left(n_{\text{r}},n_{\text{c}}\right)$.
Then according to (\ref{eq:RayonIntersection}) and (\ref{eq:Classif_GroundPix}),
$\left(n_{\text{r}},n_{\text{c}}\right)\notin\mathcal{Y}_{i}^{\text{g}}$.

The assumption (\ref{eq:RayonIntersection}) is instrumental in the
exploration process to characterize parts of $\mathbb{X}_{\text{g}}$
that cannot contain any target location. The validity of this assumption
depends on the camera orientation and on the characteristics of the
shape $\mathbb{S}_{j}^{\text{t}}\left(\mathbf{x}_{j}^{\text{t}}\right)$.
It is reasonable for many targets such as cars, trucks, provided that
they are observed from a location sufficiently above the target, with
a camera oriented towards the ground.

\section{Exploiting CVS information for set-membership estimation \label{sec:ExploitingCVS}}

Section~\ref{subsec:setEstimate} presents an approach to estimate
the location of a target~$j$ identified at time $t_{k}$ by the
CVS of UAV~$i$. Then, Section~\ref{subsec:FreeSpace} describes
how the CVS information can be used to characterize a part of $\mathbb{X}_{\text{g}}$
clear of any target. Section~\ref{subsec:Estimation-HiddenArea}
presents the characterization of areas occluded by obstacles. In the
remainder of this section, the time index $k$ is again omitted to
lighten notations.

\subsection{\textcolor{black}{Estimation of }the location of a target\label{subsec:setEstimate}}

Consider a target~$j$ such that $j\in\mathcal{D}_{i}^{\text{t}}$
with unknown location $\boldsymbol{x}_{j}^{\text{t,g}}$. UAV~$i$
has access to the bounding box $\left[\mathcal{Y}_{i,j}^{\text{t}}\right]$,
the list of pixel indexes $\mathcal{Y}_{i}^{\text{t}}$ labeled as
Target, and the depth map $\boldsymbol{\mathbf{D}}_{i}$. From these
measurements related to target~$j$, UAV~$i$ has to evaluate a
set $\mathbb{X}_{i,j}^{\text{t,m}}$ such that $\boldsymbol{x}_{j}^{\text{t,g}}\in\mathbb{X}_{i,j}^{\text{t,m}}$.

\begin{figure}
\centering{}\includegraphics[width=0.8\columnwidth]{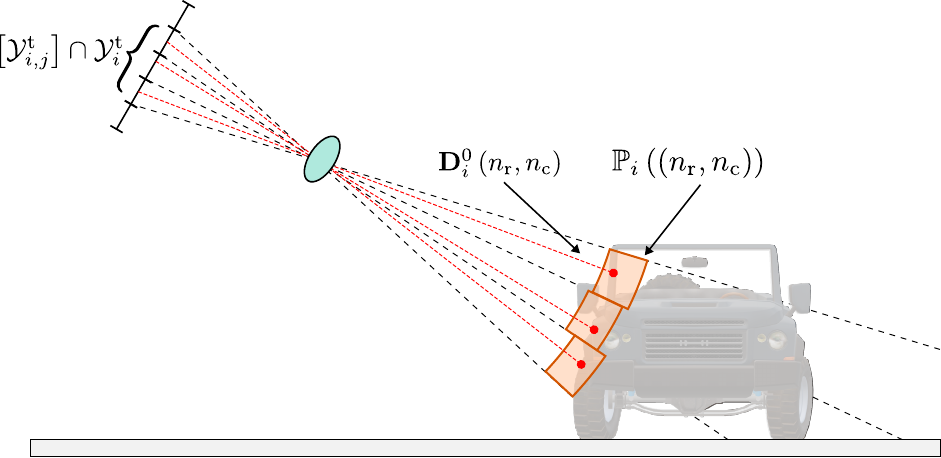}\caption{Sets $\mathbb{P}_{i}\left(\left(n_{\text{r}},n_{\text{c}}\right)\right)$
(in orange) for different $\left(n_{\text{r}},n_{\text{c}}\right)\in\left[\mathcal{Y}_{i,j}^{\text{t}}\right]$
and for a target detected and identified by UAV~$i$\label{fig:Pi}}
\end{figure}

For that purpose, a subset of $\mathbb{X}_{0}$ with a non-empty intersection
with $\mathbb{S}_{j}^{\text{t}}\left(\mathbf{x}_{j}^{\text{t}}\right)$
is first characterized. For each pixel $\left(n_{\text{r}},n_{\text{c}}\right)\in\mathbf{I}_{i}$,
consider the set 
\begin{align}
\mathbb{P}_{i}\left(\left(n_{\text{r}},n_{\text{c}}\right)\right)= & \left\{ \boldsymbol{x}\in\mathbb{F}\left(\mathbf{x}_{i}^{\text{u}}\right)\cap\mathbb{X}_{0}\mid\exists\boldsymbol{v}\in\mathcal{V}_{i}\left(n_{\text{r}},n_{\text{c}}\right),\right.\nonumber \\
 & \left.d_{\boldsymbol{v}}\left(\boldsymbol{x}_{i}^{\text{c}},\left\{ \boldsymbol{x}\right\} \right)\in\left[\mathbf{D}_{i}\right]\left(n_{\text{r}},n_{\text{c}}\right)\right\} \label{eq:Pi}
\end{align}
of all points in $\underline{\mathbb{F}}\left(\mathbf{x}_{i}^{\text{u}}\right)\cap\mathbb{X}_{0}$
that may have contributed to the illumination of $\left(n_{\text{r}},n_{\text{c}}\right)$
while being at a distance from UAV~$i$ consistent with $\mathbf{D}_{i}\left(n_{\text{r}},n_{\text{c}}\right)$,
see Figure~\ref{fig:Pi}.

\begin{figure}
\centering{}\includegraphics[width=0.8\columnwidth]{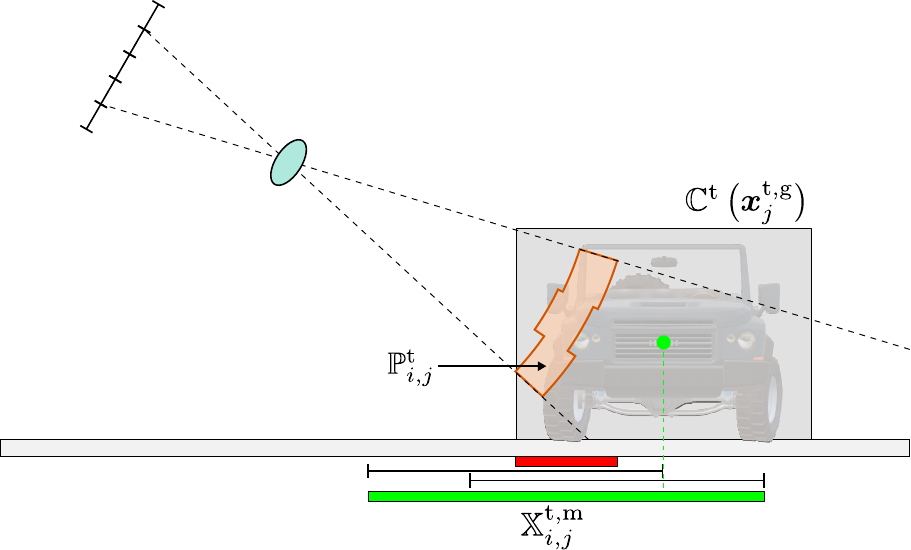}\caption{\label{fig:Detected} Estimation of the location of target~$j$:
The set $\mathbb{P}_{i,j}^{\text{t}}$ is projected on the ground
to get $\boldsymbol{p}_{\text{g}}\left(\mathbb{P}_{i,j}^{\text{t}}\right)$
(in red) ; the sets $\mathbb{D}_{\text{g}}\left(\boldsymbol{x},r^{\text{t}}\right)$
(black segments) for all $\boldsymbol{x}\in\boldsymbol{p}_{\text{g}}\left(\mathbb{P}_{i,j}^{\text{t}}\right)$
are used to build $\mathbb{X}_{i,j}^{\text{t,m}}$ (in green) ; The
green dot represents $\boldsymbol{x}_{j,k}^{\text{t}}$.}
\end{figure}

According to (\ref{eq:TargDetection_Yij}), $\mathcal{Y}_{i,j}^{\text{t}}\neq\emptyset$.
Moreover, from (\ref{eq:BoxIntersectsYt}), one has $\mathcal{Y}_{i,j}^{\text{t}}\subset\left[\mathcal{Y}_{i,j}^{\text{t}}\right]$.
Nevertheless, as indicated in Section~\ref{subsec:Targ_Detection},
only the set of pixels $\mathcal{Y}_{i}^{\text{t}}$ such that $\mathcal{Y}_{i,j}^{\text{t}}\subset\mathcal{Y}_{i}^{\text{t}}$
is available. Proposition~\ref{prop:P_cap_S_H1} states that there
exists at least one pixel $\left(n_{\text{r}},n_{\text{c}}\right)$
in $\left[\mathcal{Y}_{i,j}^{\text{t}}\right]\cap\mathcal{Y}_{i}^{\text{t}}$
for which the set $\mathbb{P}_{i}\left(\left(n_{\text{r}},n_{\text{c}}\right)\right)$
intersects $\mathbb{S}_{j}^{\text{t}}\left(\mathbf{x}_{j}^{\text{t}}\right)$.
\begin{prop}
\label{prop:P_cap_S_H1} If $j\in\mathcal{D}_{i}^{\text{t}}$, then
$\exists\left(n_{\text{r}},n_{\text{c}}\right)\in\left[\mathcal{Y}_{i,j}^{\text{t}}\right]\cap\mathcal{Y}_{i}^{\text{t}}$
such that $\mathbb{P}_{i}\left(\left(n_{\text{r}},n_{\text{c}}\right)\right)\cap\mathbb{S}_{j}^{\text{t}}\left(\mathbf{x}_{j}^{\text{t}}\right)\neq\emptyset$.
\end{prop}
\begin{IEEEproof}
If $j\in\mathcal{D}_{i}^{\text{t}}$, then $\mathcal{Y}_{i,j}^{\text{t}}\neq\emptyset$
and $\mathcal{Y}_{i,j}^{\text{t}}\subset\left[\mathcal{Y}_{i,j}^{\text{t}}\right]$.
Combining (\ref{eq:TargDetection_Yij}), (\ref{eq:BoxIntersectsYt}),
and (\ref{eq:Classif_TargPix}), $\exists\left(n_{\text{r}},n_{\text{c}}\right)\in\left[\mathcal{Y}_{i,j}^{\text{t}}\right]\cap\mathcal{Y}_{i}^{\text{t}}$
such that for all $\boldsymbol{v}\in\mathcal{V}_{i}\left(n_{\text{r}},n_{\text{c}}\right)$,
$\rho\left(\mathbf{x}_{i}^{\text{u}},\boldsymbol{v}\right)=d_{\boldsymbol{v}}\left(\boldsymbol{x}_{i}^{\text{c}},\mathbb{S}_{j}^{\text{t}}\left(\mathbf{x}_{j}^{\text{t}}\right)\right)$.
Therefore, $\exists\boldsymbol{v}\in\mathcal{V}_{i}\left(n_{\text{r}},n_{\text{c}}\right)$
and $\exists\boldsymbol{x}\in\mathbb{S}_{j}^{\text{t}}\left(\mathbf{x}_{j}^{\text{t}}\right)$
such that $d_{\boldsymbol{v}}\left(\boldsymbol{x}_{i}^{\text{c}},\left\{ \boldsymbol{x}\right\} \right)=\mathbf{D}_{i}^{0}\left(n_{\text{r}},n_{\text{c}}\right)$.
As $\left(n_{\text{r}},n_{\text{c}}\right)\in\mathcal{Y}_{i}^{\text{t}}\subset\mathcal{Y}_{i}$,
according to (\ref{eq:reliablePixels}), $\mathbf{D}_{i}\left(n_{\text{r}},n_{\text{c}}\right)/\left(1+\underline{w}\right)\leqslant d_{\max}$
and $d_{\boldsymbol{v}}\left(\boldsymbol{x}_{i}^{\text{c}},\left\{ \boldsymbol{x}\right\} \right)\leqslant d_{\max}$.
Then $\boldsymbol{x}\in\underline{\mathbb{F}}\left(\mathbf{x}_{i}^{\text{u}}\right)$.
Finally, from (\ref{eq:Pi}), we have $\boldsymbol{x}\in\mathbb{P}_{i}\left(\left(n_{\text{r}},n_{\text{c}}\right)\right)$.
So $\mathbb{P}_{i}\left(\left(n_{\text{r}},n_{\text{c}}\right)\right)\cap\mathbb{S}_{j}^{\text{t}}\left(\mathbf{x}_{j}^{\text{t}}\right)\neq\emptyset$.
\end{IEEEproof}
According to Proposition~\ref{prop:P_cap_S_H1}, a pixel $\left(n_{\text{r}},n_{\text{c}}\right)$
such that $\mathbb{P}_{i}\left(\left(n_{\text{r}},n_{\text{c}}\right)\right)\cap\mathbb{S}_{j}^{\text{t}}\left(\mathbf{x}_{j}^{\text{t}}\right)\neq\emptyset$
exists, but it is only known to belong to $\mathcal{Y}_{i}^{\text{t}}\cap\left[\mathcal{Y}_{i,j}^{\text{t}}\right]$.
To get a set intersecting $\mathbb{S}_{j}^{\text{t}}\left(\mathbf{x}_{j}^{\text{t}}\right)$,
one has then to consider the union of all sets $\mathbb{P}_{i}\left(\left(n_{\text{r}},n_{\text{c}}\right)\right)$
for the pixels $\left(n_{\text{r}},n_{\text{c}}\right)\in\left[\mathcal{Y}_{i,j}^{\text{t}}\right]\cap\mathcal{Y}_{i}^{\text{t}}$,
see Corollary~\ref{cor:P_ijk}, a direct consequence of Proposition~\ref{prop:P_cap_S_H1}.
\begin{cor}
\label{cor:P_ijk}If $j\in\mathcal{D}_{i}^{\text{t}}$, then the set
\begin{align}
\mathbb{P}_{i,j}^{\text{t}} & ={\textstyle \bigcup}_{\left(n_{\text{r}},n_{\text{c}}\right)\in\left[\mathcal{Y}_{i,j}^{\text{t}}\right]\cap\mathcal{Y}_{i}^{\text{t}}}\mathbb{P}_{i}\left(\left(n_{\text{r}},n_{\text{c}}\right)\right)\label{eq:Pijk}
\end{align}
is such that $\mathbb{P}_{i,j}^{\text{t}}\cap\mathbb{S}_{j}^{\text{t}}\left(\mathbf{x}_{j}^{\text{t}}\right)\neq\emptyset$.
\end{cor}
Again, $\mathbb{P}_{i,j}^{\text{t}}$ is only known to intersect $\mathbb{S}_{j}^{\text{t}}\left(\mathbf{x}_{j}^{\text{t}}\right)$.
If $\boldsymbol{x}\in\mathbb{P}_{i,j}^{\text{t}}\cap\mathbb{S}_{j}^{\text{t}}\left(\mathbf{x}_{j}^{\text{t}}\right)$
would be available, exploiting the fact that $\mathbb{S}_{j}^{\text{t}}\left(\mathbf{x}_{j}^{\text{t}}\right)\subset\mathbb{C}^{\text{t}}\left(\boldsymbol{x}_{j}^{\text{t,g}}\right)$,
one would have $\boldsymbol{x}_{j}^{\text{t,g}}\in\mathbb{D}_{\text{g}}\left(\boldsymbol{p}_{\text{g}}\left(\boldsymbol{x}\right),r^{\text{t}}\right)$,
see the proof of Proposition~\ref{prop:InXc-1}, below. The set estimate
$\mathbb{X}_{i,j}^{\text{t,m}}$ of $\boldsymbol{x}_{j}^{\text{t,g}}$
introduced in Proposition~\ref{prop:InXc-1} accounts for the fact
that $\boldsymbol{x}$ is only known to belong to $\mathbb{P}_{i,j}^{\text{t}}$.
\begin{prop}
\label{prop:InXc-1}If $j\in\mathcal{D}_{i}^{\text{t}}$, then the
set estimate
\begin{align}
\mathbb{X}_{i,j}^{\text{t,m}} & ={\textstyle \bigcup}_{\boldsymbol{x}\in\boldsymbol{p}_{\text{g}}\left(\mathbb{P}_{i,j}^{\text{t}}\right)}\mathbb{D}_{\text{g}}\left(\boldsymbol{x},r^{\text{t}}\right)\label{eq:setEstimate_Xijk}
\end{align}
is such that $\boldsymbol{x}_{j}^{\text{t,g}}\in\mathbb{X}_{i,j}^{\text{t,m}}$.
\end{prop}
\begin{IEEEproof}
As $j\in\mathcal{D}_{i}^{\text{t}}$, from Corollary~\ref{cor:P_ijk},
$\mathbb{P}_{i,j}^{\text{t}}\cap\mathbb{S}_{j}^{\text{t}}\left(\mathbf{x}_{j}^{\text{t}}\right)\neq\emptyset$.
Consequently, $\boldsymbol{p}_{\text{g}}\left(\mathbb{P}_{i,j}^{\text{t}}\right)\cap\boldsymbol{p}_{\text{g}}\left(\mathbb{S}_{j}^{\text{t}}\left(\mathbf{x}_{j}^{\text{t}}\right)\right)\neq\emptyset$
and there exists $\boldsymbol{x}\in\boldsymbol{p}_{\text{g}}\left(\mathbb{P}_{i,j}^{\text{t}}\right)\cap\boldsymbol{p}_{\text{g}}\left(\mathbb{S}_{j}^{\text{t}}\left(\mathbf{x}_{j}^{\text{t}}\right)\right)\subset\boldsymbol{p}_{\text{g}}\left(\mathbb{S}_{j}^{\text{t}}\left(\mathbf{x}_{j}^{\text{t}}\right)\right)$.
Since $\mathbb{S}_{j}^{\text{t}}\left(\mathbf{x}_{j}^{\text{t}}\right)\subset\mathbb{C}^{\text{t}}\left(\boldsymbol{x}_{j}^{\text{t,g}}\right)$,
$\boldsymbol{p}_{\text{g}}\left(\mathbb{S}_{j}^{\text{t}}\left(\mathbf{x}_{j}^{\text{t}}\right)\right)\subset\boldsymbol{p}_{\text{g}}\left(\mathbb{C}^{\text{t}}\left(\boldsymbol{x}_{j}^{\text{t,g}}\right)\right)$
and $\boldsymbol{x}\in\boldsymbol{p}_{\text{g}}\left(\mathbb{C}^{\text{t}}\left(\boldsymbol{x}_{j}^{\text{t,g}}\right)\right)$.
As $\boldsymbol{p}_{\text{g}}\left(\mathbb{C}^{\text{t}}\left(\boldsymbol{x}_{j}^{\text{t,g}}\right)\right)=\mathbb{D}_{\text{g}}\left(\boldsymbol{x}_{j}^{\text{t,g}},r^{\text{t}}\right)$,
one has $\left\Vert \boldsymbol{x}-\boldsymbol{x}_{j}^{\text{t,g}}\right\Vert \leqslant r^{\text{t}}$
and $\boldsymbol{x}_{j}^{\text{t,g}}\in\mathbb{D}_{\text{g}}\left(\boldsymbol{x},r^{\text{t}}\right)$.
Consequently, $\boldsymbol{x}_{j}^{\text{t,g}}\in{\textstyle \bigcup}_{\boldsymbol{x}\in\boldsymbol{p}_{\text{g}}\left(\mathbb{P}_{i,j}^{\text{t}}\right)}\mathbb{D}_{\text{g}}\left(\boldsymbol{x},r^{\text{t}}\right)$.
\end{IEEEproof}
Determining $\boldsymbol{x}\in\boldsymbol{p}_{\text{g}}\left(\mathbb{P}_{i,j}^{\text{t}}\right)\cap\boldsymbol{p}_{\text{g}}\left(\mathbb{S}_{j}^{\text{t}}\left(\mathbf{x}_{j}^{\text{t}}\right)\right)$
without knowing $\mathbb{S}_{j}^{\text{t}}\left(\mathbf{x}_{j}^{\text{t}}\right)$
is difficult. This is why the estimate $\mathbb{X}_{i,j}^{\text{t,m}}$
is defined considering all $\boldsymbol{x}\in\boldsymbol{p}_{\text{g}}\left(\mathbb{P}_{i,j}^{\text{t}}\right)$
as $\boldsymbol{p}_{\text{g}}\left(\mathbb{P}_{i,j}^{\text{t}}\right)\cap\boldsymbol{p}_{\text{g}}\left(\mathbb{S}_{j}^{\text{t}}\left(\mathbf{x}_{j}^{\text{t}}\right)\right)\subset\boldsymbol{p}_{\text{g}}\left(\mathbb{P}_{i,j}^{\text{t}}\right)$.
In practice, the estimate $\mathbb{X}_{i,j}^{\text{t,m}}$ is evaluated
as
\begin{align}
\mathbb{X}_{i,j}^{\text{t,m}} & =\mathbb{X}_{\text{g}}\cap\left(\boldsymbol{p}_{\text{g}}\left(\mathbb{P}_{i,j}^{\text{t}}\right)\oplus\mathbb{\mathbb{D}_{\text{g}}}\left(\mathbf{0},r^{\text{t}}\right)\right),\label{eq:setEstimate_Xijk_bis}
\end{align}
which, compared to (\ref{eq:setEstimate_Xijk}), only involves the
Minkovski sum of the disc $\mathbb{D}_{\text{g}}\left(\mathbf{0},r^{\text{t}}\right)$
and the projection of $\mathbb{P}_{i,j}^{\text{t}}$ on $\mathbb{X}_{\text{g}}$,
see Figure~\ref{fig:Detected}. Appendix~\ref{subsec:Appendix_CharacX}
describes the practical evaluation of $\mathbb{X}_{i,j}^{\text{t,m}}$.

\subsection{Estimation of the space free of target\label{subsec:FreeSpace}}

In what follows, using $\mathbf{L}_{i}$ and $\boldsymbol{\mathbf{D}}_{i}$,
one characterizes a subset of $\mathbb{X}_{\text{g}}$ which contains
no target location. For that purpose, pixels labeled as Ground and
as Obstacles are used respectively in Sections~\ref{subsec:FreeSpace_Ground}
and~\ref{subsec:FreeSpace_Obs}.

\subsubsection{Using pixels labeled as Ground\label{subsec:FreeSpace_Ground}}

For each pixel $\left(n_{\text{r}},n_{\text{c}}\right)$, the set
\begin{equation}
\mathbb{P}_{i}^{\text{g}}\left(\left(n_{\text{r}},n_{\text{c}}\right)\right)=\left\{ \boldsymbol{x}\in\underline{\mathbb{F}}\left(\mathbf{x}_{i}^{\text{u}}\right)\cap\mathbb{X}_{\text{g}}\mid\boldsymbol{p}_{\text{c}}\left(\mathbf{x}_{i}^{\text{u}},\boldsymbol{x}\right)\in\left(n_{\text{r}},n_{\text{c}}\right)\right\} \label{eq:Pi_G}
\end{equation}
contains all points in $\underline{\mathbb{F}}\left(\mathbf{x}_{i}^{\text{u}}\right)\cap\mathbb{X}_{\text{g}}$,
the intersection of the FoV with the ground, which image in the CCD
array belongs to $\left(n_{\text{r}},n_{\text{c}}\right)$. According
to (\ref{eq:Classif_GroundPix}), pixels in $\mathcal{Y}_{i}^{\text{g}}$,
\emph{i.e.}, labeled as Ground, are such that only points in $\mathbb{X}_{\text{g}}$
have contributed to their illumination. Consequently, the set
\begin{align}
\mathbb{P}_{i}^{\text{g}}\left(\mathcal{Y}_{i}^{\text{g}}\right)= & \left\{ \boldsymbol{x}\in\underline{\mathbb{F}}\left(\mathbf{x}_{i}^{\text{u}}\right)\cap\mathbb{X}_{\text{g}}\mid\boldsymbol{p}_{\text{c}}\left(\mathbf{x}_{i}^{\text{u}},\boldsymbol{x}\right)\in\mathcal{Y}_{i}^{\text{g}}\right\} \label{eq:Pig}
\end{align}
of points of $\mathbb{X}_{\text{g}}$ which image in the CCD array
belongs to a pixel classified as ground, cannot contain the location
of a target, see Proposition~\ref{prop:targCoM_notIn_Yg}.
\begin{prop}
\label{prop:targCoM_notIn_Yg}For all $j\in\mathcal{N}^{\text{t}}$,
$\boldsymbol{x}_{j}^{\text{t,g}}\notin\mathbb{P}_{i}^{\text{g}}\left(\mathcal{Y}_{i}^{\text{g}}\right)$.
\end{prop}
\begin{IEEEproof}
First, if $\boldsymbol{x}_{j}^{\text{t,g}}\notin\underline{\mathbb{F}}\left(\mathbf{x}_{i}^{\text{u}}\right)$,
then $\boldsymbol{x}_{j}^{\text{t,g}}\notin\mathbb{P}_{i}^{\text{g}}\left(\mathcal{Y}_{i}^{\text{g}}\right)$
by definition of $\mathbb{P}_{i}^{\text{g}}\left(\mathcal{Y}_{i}^{\text{g}}\right)$.
Assume now that $\boldsymbol{x}_{j}^{\text{t,g}}\in\underline{\mathbb{F}}\left(\mathbf{x}_{i}^{\text{u}}\right)$.
According to (\ref{eq:PixelExistence4PointsInFoV}), there exists
$\left(n_{\text{r}},n_{\text{c}}\right)\in\mathcal{N}^{\text{I}}$
such that $\boldsymbol{p}_{\text{c}}\left(\mathbf{x}_{i}^{\text{u}},\boldsymbol{x}_{j}^{\text{t,g}}\right)\in\left(n_{\text{r}},n_{\text{c}}\right)$.
Moreover, according to (\ref{eq:RayonIntersection}), there exists
a half-open segment $\left[\boldsymbol{x}_{i}^{\text{c}},\boldsymbol{x}_{j}^{\text{t,g}}\right[$
such that $\left[\boldsymbol{x}_{i}^{\text{c}},\boldsymbol{x}_{j}^{\text{t,g}}\right[\cap\mathbb{S}_{j}^{\text{t}}\left(\mathbf{x}_{j}^{\text{t}}\right)\neq\emptyset$
while $\left[\boldsymbol{x}_{i}^{\text{c}},\boldsymbol{x}_{j}^{\text{t,g}}\right[\cap\mathbb{X}_{\text{g}}=\emptyset$.
Consider the unit vector $\boldsymbol{v}\in\mathcal{V}_{i}\left(n_{\text{r}},n_{\text{c}}\right)$
such that $\boldsymbol{v}$ is colinear with $\left[\boldsymbol{x}_{i}^{\text{c}},\boldsymbol{x}_{j}^{\text{t,g}}\right[.$
Therefore, $\rho\left(\mathbf{x}_{i}^{\text{u}},\boldsymbol{v}\right)\neq d_{\boldsymbol{v}}\left(\boldsymbol{x}_{i}^{\text{c}},\mathbb{X}_{\text{g}}\right)$
and, according to (\ref{eq:Classif_GroundPix}), $\boldsymbol{p}_{\text{c}}\left(\mathbf{x}_{i}^{\text{u}},\boldsymbol{x}_{j}^{\text{t,g}}\right)\notin\mathcal{Y}_{i}^{\text{g}}$.
Consequently, from (\ref{eq:Pig}), $\boldsymbol{x}_{j}^{\text{t,g}}\notin\mathbb{P}_{i}^{\text{g}}\left(\mathcal{Y}_{i}^{\text{g}}\right)$.
\end{IEEEproof}
Appendix~\ref{subsec:Appendix_CharacP} describes an evaluation approach
for $\mathbb{P}_{i}^{\text{g}}\left(\mathcal{Y}_{i}^{\text{g}}\right)$.

\subsubsection{Using pixels labeled as Obstacle\label{subsec:FreeSpace_Obs}}

UAV~$i$ is able to characterize an inner-approximation $\underline{\mathbb{X}}_{i}^{\text{o}}\subset\mathbb{X}_{\text{g}}$
of the $r^{\text{s}}$-ground neighborhood of $\bigcup_{m\in\mathcal{N}^{\text{o}}}p_{\text{g}}\left(\mathbb{S}_{m}^{\text{o}}\right)$,
the union of the projection on the ground of the shape of all obstacles.
According to (\ref{eq:SafetyDistance}), the distance between a target
location and the projection on the ground of any obstacle is at least
$r^{\text{s}}$. Consequently, the set $\underline{\mathbb{X}}_{i}^{\text{o}}$
does not contain any target location.

For pixels labeled as Obstacle, Proposition~\ref{prop:Pi_Cap_So_FS}
provides a subset of $\mathbb{X}_{0}$ intersecting some obstacle
$\mathbb{S}_{m}^{\text{o}}$.
\begin{prop}
\label{prop:Pi_Cap_So_FS}If $\left(n_{\text{r}},n_{\text{c}}\right)\in\mathcal{Y}_{i}^{\text{o}}$,
then there exists $m\in\mathcal{N}^{\text{o}}$ such that $\mathbb{P}_{i}\left(\left(n_{\text{r}},n_{\text{c}}\right)\right)\cap\mathbb{S}_{m}^{\text{o}}\neq\emptyset$.
\end{prop}
\begin{IEEEproof}
If $\left(n_{\text{r}},n_{\text{c}}\right)\in\mathcal{Y}_{i}^{\text{o}}$,
then, according to (\ref{eq:Classif_ObsPix-1}), for all $\boldsymbol{v}\in\mathcal{V}_{i}\left(n_{\text{r}},n_{\text{c}}\right)$,
there exists $m\in\mathcal{N}^{\text{o}}$ such that $\rho\left(\mathbf{x}_{i}^{\text{u}},\boldsymbol{v}\right)=d_{\boldsymbol{v}}\left(\boldsymbol{x}_{i}^{\text{c}},\mathbb{S}_{m}^{\text{o}}\right)$.
Therefore, there exists $\boldsymbol{v}\in\mathcal{V}_{i}\left(n_{\text{r}},n_{\text{c}}\right)$
and there exists $\boldsymbol{x}\in\mathbb{S}_{m}^{\text{o}}$ such
that $d_{\boldsymbol{v}}\left(\boldsymbol{x}_{i}^{\text{c}},\left\{ \boldsymbol{x}\right\} \right)=\mathbf{D}_{i}^{0}\left(n_{\text{r}},n_{\text{c}}\right)$.
Consequently, from (\ref{eq:Pi}), $\boldsymbol{x}\in\mathbb{P}_{i}\left(\left(n_{\text{r}},n_{\text{c}}\right)\right)$.
So there exists $m\in\mathcal{N}^{\text{o}}$ such that $\mathbb{P}_{i}\left(\left(n_{\text{r}},n_{\text{c}}\right)\right)\cap\mathbb{S}_{m}^{\text{o}}\neq\emptyset$.
\end{IEEEproof}
If some $\boldsymbol{x}\in\mathbb{P}_{i}\left(\left(n_{\text{r}},n_{\text{c}}\right)\right)\cap\mathbb{S}_{m}^{\text{o}}$
would be known, then, according to (\ref{eq:SafetyDistance}), one
would be able to prove that there is no target location in $\mathbb{N}_{\text{g}}\left(\left\{ \boldsymbol{x}\right\} ,r^{\text{s}}\right)$,
the $r^{\text{s}}$-ground neighborhood of $\boldsymbol{x}$. As $\mathbb{P}_{i}\left(\left(n_{\text{r}},n_{\text{c}}\right)\right)\cap\mathbb{S}_{m}^{\text{o}}\subset\mathbb{P}_{i}\left(\left(n_{\text{r}},n_{\text{c}}\right)\right)$,
one considers the set
\begin{equation}
\mathbb{S}^{\text{o}}\left(\left(n_{\text{r}},n_{\text{c}}\right),r^{\text{s}}\right)={\textstyle \bigcap}_{\boldsymbol{x}\in\mathbb{P}_{i}\left(\left(n_{\text{r}},n_{\text{c}}\right)\right)}\mathbb{N}_{\text{g}}\left(\left\{ \boldsymbol{x}\right\} ,r^{\text{s}}\right).\label{eq:So_ExclusionSpace}
\end{equation}
defined as the intersections of all $r^{\text{s}}$-ground neighborhood
of $\boldsymbol{x}\in\mathbb{P}_{i}\left(\left(n_{\text{r}},n_{\text{c}}\right)\right)$.
The following proposition states that $\mathbb{S}^{\text{o}}\left(\left(n_{\text{r}},n_{\text{c}}\right),r^{\text{s}}\right)$
is an inner-approximation of the $r^{\text{s}}$-ground neighborhood
of an obstacle.
\begin{prop}
\label{prop:InteriorApprox_ObsNeighbor}If $\left(n_{\text{r}},n_{\text{c}}\right)\in\mathcal{Y}_{i}^{\text{o}}$,
then there exists $m\in\mathcal{N}^{\text{o}}$, such that
\begin{equation}
\mathbb{S}^{\text{o}}\left(\left(n_{r},n_{c}\right),r^{\text{s}}\right)\subset\mathbb{N}_{\text{g}}\left(\mathbb{S}_{m}^{\text{o}},r^{\text{s}}\right).\label{eq:S0subset}
\end{equation}
\end{prop}
\begin{IEEEproof}
If $\left(n_{\text{r}},n_{\text{c}}\right)\in\mathcal{Y}_{i}^{\text{o}}$,
then, according to Proposition~\ref{prop:Pi_Cap_So_FS}, there exists
$m\in\mathcal{N}^{\text{o}}$ such that $\mathbb{P}_{i}\left(\left(n_{\text{r}},n_{\text{c}}\right)\right)\cap\mathbb{S}_{m}^{\text{o}}\neq\emptyset$.
As $\mathbb{P}_{i}\left(\left(n_{\text{r}},n_{\text{c}}\right)\right)\cap\mathbb{S}_{m}^{\text{o}}\subset\mathbb{P}_{i}\left(\left(n_{\text{r}},n_{\text{c}}\right)\right)$,
one has
\begin{equation}
\bigcap_{\boldsymbol{x}\in\mathbb{P}_{i}\left(\left(n_{\text{r}},n_{\text{c}}\right)\right)}\mathbb{N}_{\text{g}}\left(\left\{ \boldsymbol{x}\right\} ,r^{\text{s}}\right)\subset\bigcap_{\boldsymbol{x}\in\mathbb{P}_{i}\left(\left(n_{\text{r}},n_{\text{c}}\right)\right)\cap\mathbb{S}_{m}^{\text{o}}}\mathbb{N}_{\text{g}}\left(\left\{ \boldsymbol{x}\right\} ,r^{\text{s}}\right).
\end{equation}
Moreover, as $\mathbb{P}_{i}\left(\left(n_{\text{r}},n_{\text{c}}\right)\right)\cap\mathbb{S}_{m}^{\text{o}}\subset\mathbb{S}_{m}^{\text{o}}$,{\small{}
\begin{align}
\bigcap_{\boldsymbol{x}\in\mathbb{P}_{i}\left(\left(n_{\text{r}},n_{\text{c}}\right)\right)\cap\mathbb{S}_{m}^{\text{o}}}\mathbb{N}_{\text{g}}\left(\left\{ \boldsymbol{x}\right\} ,r^{\text{s}}\right) & \subset\bigcup_{\boldsymbol{x}\in\mathbb{S}_{m}^{\text{o}}}\mathbb{N}_{\text{g}}\left(\left\{ \boldsymbol{x}\right\} ,r^{\text{s}}\right)=\mathbb{N}_{\text{g}}\left(\mathbb{S}_{m}^{\text{o}},r^{\text{s}}\right).
\end{align}
}Combining the two previous inclusions, one gets (\ref{eq:S0subset}).
\end{IEEEproof}
An inner-approximation of the $r^{\text{s}}$-ground neighborhood
of all the obstacles located within the FoV of UAV~$i$ can then
be obtained by characterizing the union of the sets $\mathbb{S}^{\text{o}}\left(\left(n_{\text{r}},n_{\text{c}}\right),r^{\text{s}}\right)$
for all pixels~$\left(n_{\text{r}},n_{\text{c}}\right)\in\mathcal{Y}_{i}^{\text{o}}$.
Then using Proposition~\ref{prop:NoTargetObstacles}, UAV~$i$ is
able to evaluate a set $\underline{\mathbb{X}}_{i}^{\text{o}}$ which
cannot contain any target location.
\begin{prop}
\label{prop:NoTargetObstacles}If $\mathcal{Y}_{i}^{\text{o}}\neq\emptyset$,
then the set
\begin{equation}
\underline{\mathbb{X}}_{i}^{\text{o}}={\textstyle \bigcup}_{\left(n_{r},n_{c}\right)\in\mathcal{Y}_{i}^{\text{o}}}\mathbb{S}^{\text{o}}\left(\left(n_{\text{r}},n_{\text{c}}\right),r^{\text{s}}\right)\label{eq:Xo_ExclusionSpace}
\end{equation}
is such that $\boldsymbol{x}_{j}^{\text{t,g}}\notin\underline{\mathbb{X}}_{i}^{\text{o}}$,
for all $j\in\mathcal{N}^{\text{t}}$.
\end{prop}
\begin{IEEEproof}
Using Proposition~\ref{prop:InteriorApprox_ObsNeighbor}, for each
$\left(n_{\text{r}},n_{\text{c}}\right)\in\mathcal{Y}_{i}^{\text{o}}$,
there exists $m\in\mathcal{N}^{\text{o}}$ such that $\mathbb{S}^{\text{o}}\left(\left(n_{\text{r}},n_{\text{c}}\right),r^{\text{s}}\right)\subset\mathbb{N}_{\text{g}}\left(\mathbb{S}_{m}^{\text{o}},r^{\text{s}}\right).$
Consequently,
\begin{equation}
{\textstyle \bigcup}_{\left(n_{r},n_{c}\right)\in\mathcal{Y}_{i}^{\text{o}}}\mathbb{S}^{\text{o}}\left(\left(n_{r},n_{c}\right),r^{\text{s}}\right)\subset{\textstyle \bigcup}_{m\in\mathcal{N}^{\text{o}}}\mathbb{N}_{\text{g}}\left(\mathbb{S}_{m}^{\text{o}},r^{\text{s}}\right).
\end{equation}
Consequently $\underline{\mathbb{X}}_{i}^{\text{o}}\subset\bigcup_{m\in\mathcal{N}^{\text{o}}}\mathbb{N}_{\text{g}}\left(\mathbb{S}_{m}^{\text{o}},r^{\text{s}}\right)$.
According to Assumption~\ref{eq:SafetyDistance}, we have
\begin{align}
\forall j\in\mathcal{N}^{\text{t}},\forall m\in\mathcal{N}^{\text{o}}, & \boldsymbol{x}_{j}^{\text{t,g}}\notin\mathbb{N}_{\text{g}}\left(\mathbb{S}_{m}^{\text{o}},r^{\text{s}}\right)\nonumber \\
\Leftrightarrow\forall j\in\mathcal{N}^{\text{t}}, & \boldsymbol{x}_{j}^{\text{t,g}}\notin{\textstyle \bigcup}_{m\in\mathcal{N}^{\text{o}}}\mathbb{N}_{\text{g}}\left(\mathbb{S}_{m}^{\text{o}},r^{\text{s}}\right)
\end{align}
therefore $\forall j\in\mathcal{N}^{\text{t}},\boldsymbol{x}_{j}^{\text{t,g}}\notin\underline{\mathbb{X}}_{i}^{\text{o}}.$
\end{IEEEproof}
Appendix~\ref{subsec:Appendix_CharacX} describes the practical evaluation
of $\underline{\mathbb{X}}_{i}^{\text{o}}$.

\subsection{Estimation of the hidden area\label{subsec:Estimation-HiddenArea}}

Some parts of $\mathbb{F}\left(\mathbf{x}_{i}^{\text{u}}\right)\cap\mathbb{X}_{\text{g}}$
cannot be seen by UAV~$i$ when they are hidden by an obstacle, a
target, or a UAV. The pixels labeled either as Obstacle, Target, or
Unknown are used to characterize the hidden parts of $\mathbb{F}\left(\mathbf{x}_{i}^{\text{u}}\right)\cap\mathbb{X}_{\text{g}}$,
denoted as
\begin{align}
\mathbb{H}_{i}^{\text{g}} & =\mathbb{P}_{i}^{\text{g}}\left(\mathcal{Y}_{i}^{\text{o}}\cup\mathcal{Y}_{i}^{\text{t}}\cup\mathcal{Y}_{i}^{\text{n}}\right)\\
 & ={\textstyle \bigcup}_{\left(n_{\text{r}},n_{\text{c}}\right)\in\mathcal{Y}_{i}^{\text{o}}\cup\mathcal{Y}_{i}^{\text{t}}\cup\mathcal{Y}_{i}^{\text{n}}}\mathbb{P}_{i}^{\text{g}}\left(\left(n_{\text{r}},n_{\text{c}}\right)\right).
\end{align}
$\mathbb{H}_{i}^{\text{g}}$ represent the part of the ground that
is hidden by obstacles or targets, or for which no reliable enough
information is available. $\mathbb{H}_{i}^{\text{g}}$ will be useful
in the UAV exploration process.

Appendix~\ref{subsec:Appendix_CharacP} describes the practical evaluation
of $\mathbb{H}_{i}^{\text{g}}$.

\section{Estimation algorithm\label{sec:Prediction_Correction}}

This section describes the way a distributed set-membership target
location estimator exploits the information provided by the CVS. It
is adapted from \cite{ibenthal_bounded-error_2021}. Only the parts
related to the exploitation of the CVS are detailed here. The set
estimates are initialized at time $t_{0}$ as $\mathcal{L}_{i,0}^{\text{t}}=\emptyset$,
$\mathcal{X}_{i,0}^{\text{t}}=\emptyset$, $\overline{\mathbb{X}}_{i,0}^{\text{t}}=\mathbb{X}_{\text{g}}$,
and $\mathbb{X}_{i,0}^{\text{o}}=\emptyset$.

\subsection{Prediction step\label{subsec:Prediction-step}}

At time $t_{k}$, the predicted set $\mathbb{X}_{i,j,k\mid k-1}^{\text{t}}$
of possible future locations of identified target $j\in\mathcal{L}_{i,k}^{\text{t}}$
is evaluated using the set of previous locations $\mathbb{X}_{i,j,k-1}^{\text{t}}$
, the target dynamics (\ref{eq:DynTarget}), and the known bounds
of its control input
\begin{align}
\mathbb{X}_{i,j,k\mid k-1}^{\text{t}} & =\mathbf{f}^{\text{t}}\left(\mathbb{X}_{i,j,k-1}^{\text{t}},\left[\boldsymbol{v}^{\text{t}}\right]\right)\cap\mathbb{X}_{\text{g}}.\label{eq:PredEstimate}
\end{align}
In the same way, the predicted set $\overline{\mathbb{X}}_{i,k|k-1}^{\text{t}}$
for all possible locations of targets still to be identified is
\begin{align}
\overline{\mathbb{X}}_{i,k\mid k-1}^{\text{t}} & =\mathbf{f}^{\text{t}}\left(\overline{\mathbb{X}}_{i,k-1}^{\text{t}},\left[\boldsymbol{v}^{\text{t}}\right]\right)\cap\mathbb{X}_{\text{g}}.\label{eq:PredBarEstimate}
\end{align}
Obstacles being statics, the predicted set $\mathbb{X}_{i,k|k-1}^{\text{o}}$
is
\begin{align}
\mathbb{X}_{i,k|k-1}^{\text{o}} & =\mathbb{X}_{i,k-1}^{\text{o}}.
\end{align}

\subsection{Updates from measurements\label{subsec:Correc_Measure}}

The CVS of UAV~$i$ provides $\mathbf{I}_{i,k}$, $\mathbf{L}_{i,k}$,
$\boldsymbol{\mathbf{D}}_{i,k}$, $\mathcal{D}_{i,k}^{\text{t}}$,
and $\mathcal{B}_{i,k}^{\text{t}}$. Then $\mathcal{Y}_{i,k}^{\text{t}}$,
$\mathcal{Y}_{i,k}^{\text{g}}$, and $\mathcal{Y}_{i,k}^{\text{o}}$
are directly obtained from $\mathbf{L}_{i,k}$ and $\boldsymbol{\mathbf{D}}_{i,k}$,
see Section~\ref{subsec:Classifier}. Moreover, $\mathbb{X}_{i,j,k}^{\text{t,m}}$
is deduced from $\mathcal{Y}_{i,k}^{\text{t}}$ and $\left[\mathcal{Y}_{i,j,k}^{\text{t}}\right]$
for all $j\in\mathcal{D}_{i,k}^{\text{t}}$ using (\ref{eq:Pijk})
and (\ref{eq:setEstimate_Xijk_bis}).

First, using the set of newly identified targets $\mathcal{D}_{i,k}^{\text{t}}$,
the list of all identified targets becomes $\mathcal{L}_{i,k\mid k}^{\text{t}}=\mathcal{L}_{i,k-1}^{\text{t}}\cup\mathcal{D}_{i,k}^{\text{t}}.$

According to Sections~\ref{subsec:FreeSpace_Ground} and \ref{subsec:FreeSpace_Obs},
the set $\mathbb{P}_{i,k}^{\text{g}}\left(\mathcal{Y}_{i,k}^{\text{g}}\right)\cup\mathbb{X}_{i,k\mid k}^{\text{o}}$
is proved to be free of target. Therefore, the set $\overline{\mathbb{X}}_{i,k\mid k-1}^{\text{t}}$
is updated as 
\begin{equation}
\overline{\mathbb{X}}_{i,k\mid k}^{\text{t}}=\overline{\mathbb{X}}_{i,k\mid k-1}^{\text{t}}\setminus\left(\mathbb{P}_{i,k}^{\text{g}}\left(\mathcal{Y}_{i,k}^{\text{g}}\right)\cup\mathbb{X}_{i,k\mid k}^{\text{o}}\right).
\end{equation}
Moreover, for each previously identified target $j\in\mathcal{L}_{i,k-1}^{\text{t}}\setminus\mathcal{D}_{i,k}^{\text{t}}$
which is not identified at time $t_{k}$, $\mathbb{X}_{i,j,k\mid k-1}^{\text{t}}$
is updated as
\begin{equation}
\mathbb{X}_{i,j,k\mid k}^{\text{t}}=\mathbb{X}_{i,j,k\mid k-1}^{\text{t}}\setminus\left(\mathbb{P}_{i,k}^{\text{g}}\left(\mathcal{Y}_{i,k}^{\text{g}}\right)\cup\mathbb{X}_{i,k\mid k}^{\text{o}}\right).
\end{equation}

If target~$j$ is identified at time $t_{k}$, \textit{i.e.}, $j\in\mathcal{D}_{i,k}^{\text{t}}$,
then $\boldsymbol{x}_{j,k}^{\text{t,g}}\in\mathbb{X}_{i,j,k}^{\text{t,m}}$,
see Section~\ref{subsec:setEstimate}. Moreover, if $j\in\mathcal{L}_{i,k-1}^{\text{t}}$,
one has also $\boldsymbol{x}_{j,k}^{\text{t,g}}\in\mathbb{X}_{i,j,k\mid k-1}^{\text{t}}$
else $\boldsymbol{x}_{j,k}^{\text{t,g}}\in\overline{\mathbb{X}}_{i,k\mid k-1}^{\text{t}}$.
Therefore, for each target~$j\in\mathcal{D}_{i,k}^{\text{t}}$, one
has, if $j\in\mathcal{L}_{i,k-1}^{\text{t}}${\small{}
\begin{equation}
\mathbb{X}_{i,j,k\mid k}^{\text{t}}=\left(\mathbb{X}_{i,j,k\mid k-1}^{\text{t}}\cap\mathbb{X}_{i,j,k}^{\text{t,m}}\right)\setminus\left(\mathbb{P}_{i,k}^{\text{g}}\left(\mathcal{Y}_{i,k}^{\text{g}}\right)\cup\mathbb{X}_{i,k\mid k}^{\text{o}}\right),
\end{equation}
and if }$j\notin\mathcal{L}_{i,k-1}^{\text{t}}$
\[
\mathbb{X}_{i,j,k\mid k}^{\text{t}}=\left(\overline{\mathbb{X}}_{i,k\mid k-1}^{\text{t}}\cap\mathbb{X}_{i,j,k}^{\text{t,m}}\right)\setminus\left(\mathbb{P}_{i,k}^{\text{g}}\left(\mathcal{Y}_{i,k}^{\text{g}}\right)\cup\mathbb{X}_{i,k\mid k}^{\text{o}}\right).
\]

Finally, the $r_{s}$-neighborhoods of all obstacles is updated as
\begin{equation}
\mathbb{X}_{i,k\mid k}^{\text{o}}=\mathbb{X}_{i,k-1}^{\text{o}}\cup\underline{\mathbb{X}}_{i,k}^{\text{o}},
\end{equation}
see Section~\ref{subsec:FreeSpace_Obs}.

\subsection{Update after communication with neighbors\label{subsec:Correc-comm}}

Each UAV~$\ell$, once it has taken into account its own measurements,
broadcasts the list of identified targets $\mathcal{L}_{\ell,k|k}^{\text{t}}$,
the $r_{s}$-neighborhoods of obstacles $\mathbb{X}_{\ell,k\mid k}^{\text{o}}$,
the set estimates $\mathbb{X}_{\ell,j,k\mid k}^{\text{t}}$, $j\in\mathcal{L}_{\ell,k|k}^{\text{t}}$
and the set of possible locations of targets still to be identified
$\overline{\mathbb{X}}_{\ell,k|k}^{\text{t}}$.

Accounting for the information received from its neighbors, UAV~$i$
updates the list of identified targets as $\mathcal{L}_{i,k}^{\text{t}}=\bigcup_{\ell\in\mathcal{N}_{i,k}}\mathcal{L}_{\ell,k|k}^{\text{t}}$,
the $r_{s}$-neighborhoods of obstacles $\mathbb{X}_{i,k}^{\text{o}}=\bigcup_{\ell\in\mathcal{N}_{i,k}}\mathbb{X}_{\ell,k\mid k}^{\text{o}}$,
and the set of possible locations of targets still to be identified
$\overline{\mathbb{X}}_{i,k}^{\text{t}}=\bigcap_{\ell\in\mathcal{N}_{i,k}}\overline{\mathbb{X}}_{\ell,k|k}^{\text{t}}.$
Then, if $j\in\mathcal{L}_{i,k}^{\text{t}}$, one has 
\begin{equation}
\mathbb{X}_{i,j,k}^{\text{t}}=\bigcap_{\underset{\text{ st }j\in\mathcal{L}_{\ell,k|k}^{\text{t}}}{\ell\in\mathcal{N}_{i,k}}}\mathbb{X}_{\ell,j,k\mid k}^{\text{t}}\cap\bigcap_{\underset{\text{ st }j\notin\mathcal{L}_{\ell,k|k}^{\text{t}}}{\ell\in\mathcal{N}_{i,k}}}\overline{\mathbb{X}}_{\ell,k\mid k}^{\text{t}}.\label{eq:SetEstimatejFusion}
\end{equation}
to account (\emph{i}) for the estimate
$\mathbb{X}_{\ell,j,k\mid k}^{\text{t}}$ of the location of target
$j$ made by the neighbors of UAV $i$ which have already identified
target~$j$ and (\emph{ii}) for the fact that $\boldsymbol{x}_{j,k}^{\text{t}}\in\overline{\mathbb{X}}_{\ell,k\mid k}^{\text{t}}$
for all neighbors of UAV~$i$ which have not yet identified target~$j$.

\section{Evaluation of UAV trajectories\label{sec:MPC}}

The trajectory of UAV~$i$ is designed so as to a minimize the estimation
uncertainty of target locations (\ref{eq:TargLocUncertaintyEstimation}).
For that purpose, a distributed MPC approach \cite{christofides_distributed_2013}
is used, based on that presented in \cite{ibenthal_bounded-error_2021}.
At time $t_{k}$, the trajectory of UAV~$\mathit{i}$ is determined
over a prediction horizon $h$ by evaluating a sequence of control
inputs $\mathbf{u}_{i,k:h}^{\text{P}}=(\mathbf{u}_{i,k}^{\text{P}},\dots,\mathbf{u}_{i,k+h-1}^{\text{P}})\in\mathbb{U}\times\dots\times\mathbb{U}$,
from which a sequence of states $\mathbf{x}_{i,k:h\mid k}^{\text{u},\text{P}}=(\mathbf{x}_{i,k+1\mid k}^{\text{\text{u},\text{P}}},\dots,\mathbf{x}_{i,k+h\mid k}^{\text{u},\text{P}})$
is deduced using (\ref{eq:UAV_Dyn}), so as to minimize the criterion
\begin{align}
J_{0}\left(\mathbf{u}_{i,k:h}^{\text{P}}\right) & =\Phi\left(\mathcal{X}_{i,k+h\mid k}^{\text{t},\text{P}},\overline{\mathbb{X}}_{i,k+h\mid k}^{\text{t},\text{P}}\right).\label{eq:CritVolume}
\end{align}

The sequence of predicted states $\mathbf{x}_{i,k:h\mid k}^{\text{u},\text{P}}$
is then fed as a desired state target to the low-level controller
for UAV~$i$, responsible of evaluating high-frequency control inputs.
In (\ref{eq:CritVolume}), $\mathcal{X}_{i,k+h\mid k}^{\text{t},\text{P}}$
is the list of predicted estimates of the target locations $\mathbb{X}_{i,j,k+h\mid k}^{\text{t},\text{P}}$
for all $j\in\mathcal{L}_{i,k}^{\text{t}}$, and $\overline{\mathbb{X}}_{i,k+h\mid k}^{\text{t},\text{P}}$
is the predicted set of possible locations of targets still to be
identified, both evaluated at time $t_{k+h}$ using the information
available at time $t_{k}$. See \cite{ibenthal_bounded-error_2021}
for more details.

In some cases, $J_{0}(\mathbf{u}_{i,k:h}^{\text{P}})$ remains constant
whatever $\mathbf{u}_{i,k:h}^{\text{P}}\in\mathbb{U}^{h}$. This may
occur once large parts of the environment have been explored. A second
criterion is introduced 
\begin{align}
J_{1}\left(\mathbf{u}_{i,k:h}^{\text{P}}\right)= & d\left(c\left(\mathbb{F}\left(\mathbf{x}_{i,k+h\mid k}^{\text{u,P}}\right)\cap\mathbb{X}_{\text{g}}\right),\right.\nonumber \\
 & \hspace{-1cm}\left.\left(\overline{\mathbb{X}}_{i,k+h\mid k}^{\text{t,P}}\cup{\textstyle \bigcup}_{j\in\mathcal{L}_{i,k}^{\text{t}}}\mathbb{X}_{i,j,k+h\mid k}^{\text{t,P}}\right)\setminus\mathbb{H}_{i,k}^{\text{g}}\right),\label{eq:CritDistance}
\end{align}
where $c\left(\mathbb{X}\right)$ is the barycenter of the set $\mathbb{X}$.
The idea is to drive the FoV of UAV~$i$ to the closest predicted
set estimate related to an identified target or to the part of the
search area where targets may have still to be identified. To account
for possible occlusions by obstacles, we assume that the part $\mathbb{H}_{i,k}^{\text{g}}$
of the ground hidden at time $t_{k}$ does not evolve during the $h$
next time steps.

The resulting criterion combining (\ref{eq:CritVolume}) and (\ref{eq:CritDistance})
is
\begin{equation}
J\left(\mathbf{u}_{i,k:h}^{\text{P}}\right)=J_{0}\left(\mathbf{u}_{i,k:h}^{\text{P}}\right)+\lambda J_{1}\left(\mathbf{u}_{i,k:h}^{\text{P}}\right).\label{eq:Crit}
\end{equation}
where $\lambda$ trades off the direct reduction of the size of the
set estimates via $J_{0}(\mathbf{u}_{i,k:h}^{\text{P}})$ and future
reductions via $J_{1}(\mathbf{u}_{i,k:h}^{\text{P}})$.

Compared to \cite{ibenthal_bounded-error_2021}, $J_{1}(\mathbf{u}_{i,k:h}^{\text{P}})$
and the characterization of $\mathcal{X}_{i,k+h\mid k}^{\text{t},\text{P}}$
and $\overline{\mathbb{X}}_{i,k+h\mid k}^{\text{t},\text{P}}$ take
into account the occluded part of the ground $\mathbb{H}_{i,k}^{\text{g}}$
that UAV~$i$ has not observed at time $t_{k}$. The aim is to drive
the UAVs toward areas which are less likely to be occluded by obstacles.

\section{Simulations\label{sec:Simulations}}

Simulations results have been obtained using Webots \cite{webots_httpwwwcyberboticscom_nodate}
to generate UAV and target displacements as well as the measurements
collected by the UAVs. The set estimators and trajectory design algorithms
are implemented on Matlab.

\subsection{Simulation conditions}

\begin{figure}
\begin{centering}
\includegraphics[width=0.8\columnwidth]{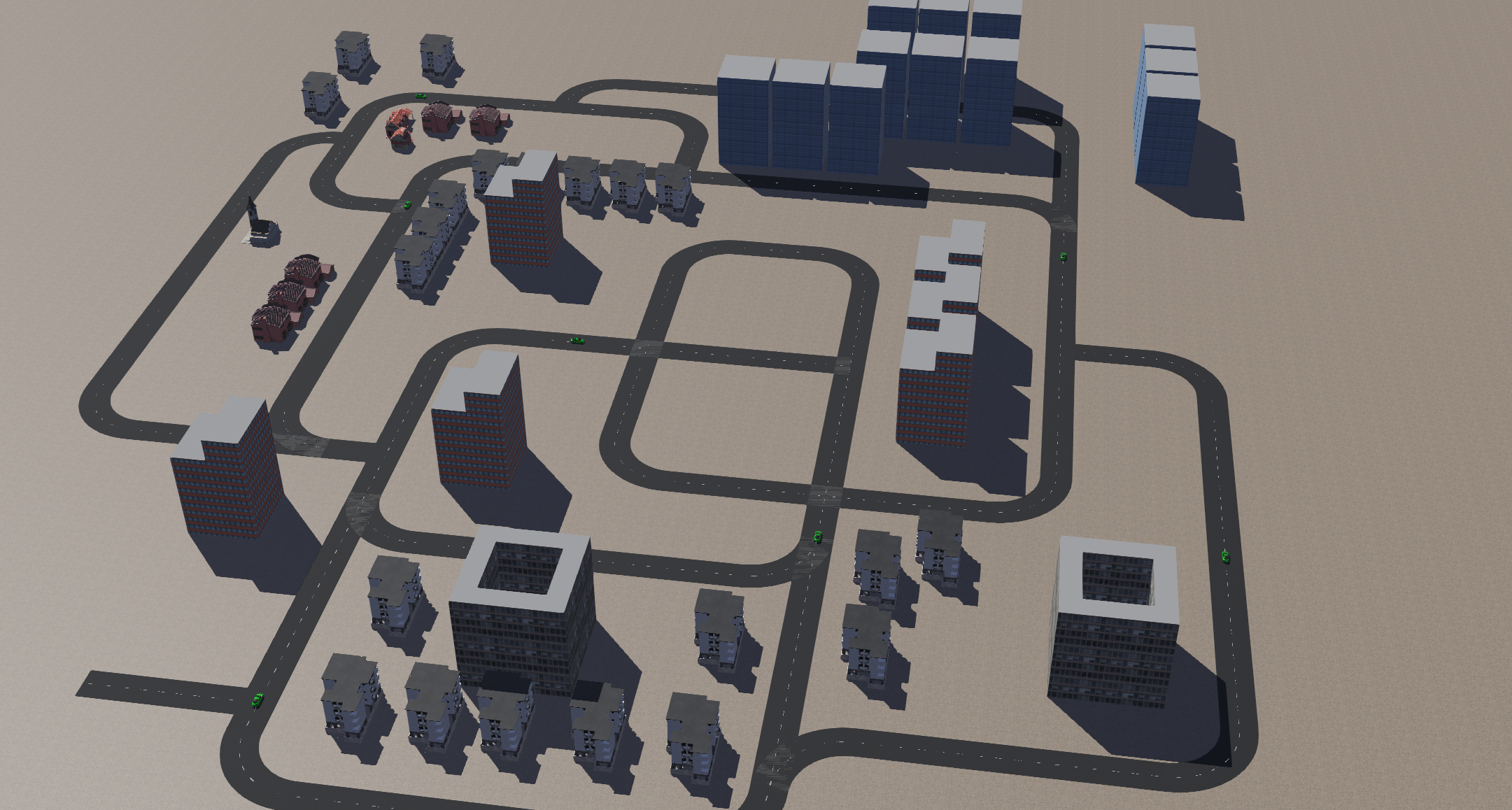}
\par\end{centering}
\caption{\label{fig:Simulated-world}Simulated urban environment in Webots}
\end{figure}

The RoI is a simplified urban environment with different types of
buildings with a height less than $50$~m, see Figure~\ref{fig:Simulated-world}.
One has $\mathbb{X}_{\text{0}}=\left[-250\,\text{m},250\,\text{m}\right]\times\left[-250\,\text{m},250\,\text{m}\right]\times\mathbb{R}^{+}$.
The ground area occupied by building represents $5$~\% of the total
area of $\mathbb{X}_{\text{g}}$. The RoI contains $N^{\text{t}}=8$
targets (cars of the same shape).

Each car has the following dimensions $4.6\ \text{m}\times1.8\ \text{m}\times1.5\ \text{m}$,
and can be included in a cylinder $\mathbb{C}^{\text{t}}$ of radius
$r^{\text{t}}=2.5$~m and height $h^{\text{t}}=2$~m. The safety
distance is taken as $r_{\text{s}}^{\text{to}}=3$~m. The state $\mathbf{x}_{j,k}^{\text{t}}=\left(x_{j,1,k}^{\text{t}},x_{j,2,k}^{\text{t}},x_{j,3,k}^{\text{t}}\right)$
of target~$j\in\mathcal{N^{\text{t}}}$ consists of the coordinates
$\left(x_{j,1,k}^{\text{t}},x_{j,2,k}^{\text{t}}\right)$ of the projection
on $\mathbb{X}_{\text{g}}$ of its center of mass, and of its heading
angle $x_{j,3,k}^{\text{t}}$. Its dynamic (\ref{eq:DynTarget}) is
\begin{equation}
\begin{pmatrix}x_{j,1,k+1}^{\text{t}}\\
x_{j,2,k+1}^{\text{t}}\\
x_{j,3,k+1}^{\text{t}}
\end{pmatrix}=\begin{pmatrix}x_{j,1,k}^{\text{t}}+Tv_{j,1}^{\text{t}}\cos\left(x_{j,3,k}^{\text{t}}\right)\\
x_{j,2,k}^{\text{t}}+Tv_{j,1}^{\text{t}}\sin\left(x_{j,3,k}^{\text{t}}\right)\\
x_{j,3,k}^{\text{t}}+Tv_{j,2,k}^{\text{t}}
\end{pmatrix},
\end{equation}
where $v_{j,1}^{\text{t}}$ is the constant target speed and $v_{j,2,k}^{\text{t}}$
is the target turn rate. At the start of the simulation, for each
target $j\in\mathcal{N}^{\text{t}}$, $v_{j,1}^{\text{t}}$ is randomly
chosen in the interval $\left[0,v_{\max}\right]$. Here, only $v_{\max}=1$~m/s
is known to the UAVs. At each time instant, $v_{j,2,k}^{\text{t}}$
is constrained to maintain the car on the roads. The UAVs do not exploit
any knowledge related to the roads during the search.

\begin{figure}
\begin{centering}
\includegraphics[width=1\columnwidth]{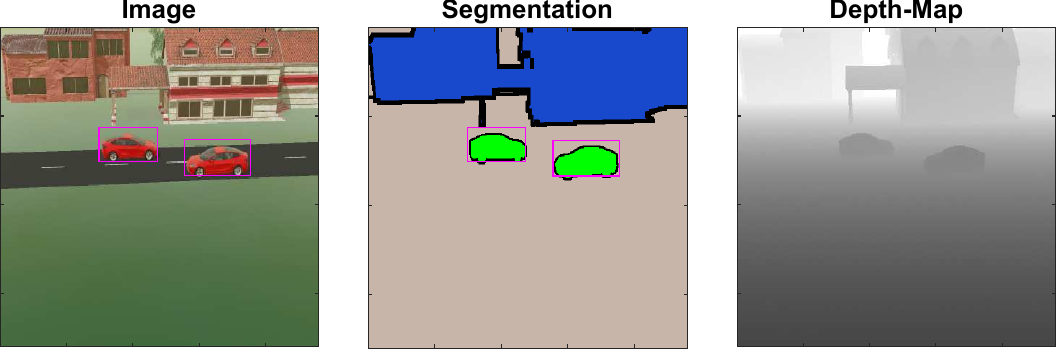}
\par\end{centering}
\caption{\label{fig:CVS_3FIG}CVS information provided to each UAV by the sensors
simulated by Webots}
\end{figure}

A fleet of $4$ identical quadcopters is considered (the DJI Mavic2Pro
model in Webots). These UAVs are assumed to be equipped with a camera
of resolution $N_{\text{r}}\times N_{\text{c}}=360\times480$ pixels,
aperture angle of $\pi/4$~rad, and orientation in the UAV body frame
of $\theta=\pi/6$~rad. A depth map is provided by the \textit{range
finder} of Webots with the same resolution as the camera. Its maximal
measurement range is set to $d_{\max}=300\,\text{m}$ and the bounds
of the noise $w$ in (\ref{eq:D=00003DD0(1+w)}) are $\left[\underline{w},\overline{w}\right]=\left[-1\%,1\%\right]$.
For example, if the distance $\mathbf{D}_{i}^{0}\left(n_{\text{r}},n_{\text{c}}\right)$
to an object is $200$~m, then the depth-map measurement $\mathbf{D}_{i}\left(n_{\text{r}},n_{\text{c}}\right)$
is obtained by randomly selecting $w\in\left[\underline{w},\overline{w}\right]=\left[-1\%,1\%\right]$,
to get, for example, $\mathbf{D}_{i}\left(n_{\text{r}},n_{\text{c}}\right)=201$~m.
Then, using (\ref{eq:DepthMap_Interval}), one gets $\left[\mathbf{D}_{i}\right]\left(n_{\text{r}},n_{\text{c}}\right)=\left[199.01\ \text{m},203.03\ \text{m}\right]$.
Figure~\ref{fig:CVS_3FIG} shows an example of CVS information provided
to each UAV.

CVS provides information with a period $T=0.5$~s. The set estimates
are updated with the same period using the algorithms presented in
Sections~\ref{sec:ExploitingCVS} and \ref{sec:Prediction_Correction}.
The trajectories of UAVs are updated with a period of $T^{\text{MPC}}=3$~s
due to the flight time constants of the quadcopters. This choice ensures
that each UAV reaches the desired target state designed by the MPC
approach. Furthermore, no restriction is considered on the communication
range between UAVs.

All UAVs are launched from the same location outside the RoI. For
the UAV trajectory design in the MPC approach, a simple dynamic model
with constant altitude and speed $v^{\text{u}}$, as well as null
roll and pitch angles is considered{\small{}
\begin{equation}
\left(\begin{array}{c}
x_{i,1,k+1}^{\text{u}}\\
x_{i,2,k+1}^{\text{u}}\\
x_{i,3,k+1}^{\text{u}}\\
x_{i,4,k+1}^{\text{u}}
\end{array}\right)=\left(\begin{array}{c}
x_{i,1,k}^{\text{u}}+Tv^{\text{\ensuremath{\text{u}}}}\cos\left(x_{i,4,k}^{\text{u}}+u_{i,k}\right)\\
x_{i,2,k}^{\text{u}}+Tv^{\text{\ensuremath{\text{u}}}}\sin\left(x_{i,4,k}^{\text{u}}+u_{i,k}\right)\\
x_{i,3,k+1}^{\text{u}}\\
x_{i,4,k}^{\text{u}}+u_{i,k}
\end{array}\right).
\end{equation}
}The control $u_{i,k}$ is a yaw angle increment taken in $\mathbb{U}=\left\{ -\pi/18,-\pi/36,0,\pi/36,\pi/18\right\} $.
The prediction horizon of the MPC is set to $h=12$. Nevertheless,
only $u_{i,k}$ and $u_{i,k+h/2}$ are chosen freely in $\mathbb{U}$,
then $u_{i,k+\tau}=u_{i,k}$ and $u_{i,k+h/2+\tau}=u_{i,k+h/2}$ for
$\tau=1,\dots,h/2-1$, \emph{i.e.}, the MPC is designed by computing
only two values of the control, one at the beginning of the MPC and
one at the middle of the prediction horizon. Moreover, to reduce computational
burden, only the predicted FoV $\underline{\mathbb{F}}\left(\mathbf{x}_{i,k+h/2\mid k}^{\text{u,P}}\right)$
and $\underline{\mathbb{F}}\left(\mathbf{x}_{i,k+h\mid k}^{\text{u,P}}\right)$
are used in the MPC approach.

At time $t_{k}$, the MPC provides $\mathbf{x}_{i,k:h}^{\text{u}}$
to a low-level PID controller designed to ensure a constant speed
module of $v^{\text{\ensuremath{\text{u}}}}=5$~m/s and a constant
fly height, higher than all obstacles and different for each UAV to
avoid collision.

\subsection{Evaluation metrics}

The performance of the propose approach is evaluated considering the
average surface of the set estimates of identified targets
\begin{equation}
\Phi_{i,k}^{\text{t}}=\frac{1}{\text{card}(\mathcal{L}_{i,k}^{\text{t}})}{\textstyle \sum}_{j\in\mathcal{L}_{i,k}^{\text{t}}}\phi\left(\mathbb{X}_{i,j,k}^{\text{t}}\right),
\end{equation}
as well as the average localization error with respect to a point
estimate taken as the barycenter $c\left(\mathbb{X}_{i,j,k}^{\text{t}}\right)$
of the set estimate
\begin{equation}
e_{i,k}^{\text{t}}=\frac{1}{\text{card}(\mathcal{L}_{i,k}^{\text{t}})}{\textstyle \sum}_{j\in\mathcal{L}_{i,k}^{\text{t}}}\left\Vert \boldsymbol{x}_{j,k}^{\text{t,g}}-c\left(\mathbb{X}_{i,j,k}^{\text{t}}\right)\right\Vert .
\end{equation}

The theoretical coverage of the FoV of the UAVs at time $k$
\begin{equation}
\Phi_{k}^{\text{FoV}}=\phi\left({\textstyle {\textstyle \bigcup}_{i\in\mathcal{N}^{\text{u}}}}\left(\mathbb{F}\left(\mathbf{x}_{i,k}^{\text{u}}\right)\cap\mathbb{X}_{\text{g}}\right)\right)
\end{equation}
represents the part of $\mathbb{X}_{\text{g}}$ observed at time $t_{k}$
by all UAVs if there is no obstacle nor targets. The part of the ground
that is actually seen and identified as ground at time $t_{k}$ is
\begin{equation}
\Phi_{k}^{\text{g}}=\phi\left({\textstyle \bigcup}_{i\in\mathcal{N}^{\text{u}}}\mathbb{P}_{i}^{\text{g}}\left(\mathcal{Y}_{i,k}^{\text{g}}\right)\right).
\end{equation}
The cumulated ground surface observed up to time $t_{k}$ is 
\begin{equation}
\Phi_{i,k}^{\text{Cg}}=\phi\left({\textstyle \bigcup_{\tau=1}^{k}}{\textstyle \bigcup}_{i\in\mathcal{N}^{\text{u}}}\mathbb{P}_{\ell}^{\text{g}}\left(\mathcal{Y}_{\ell,\tau}^{\text{g}}\right)\right).
\end{equation}
Most of these metrics may be expressed relative to the ground surface
$\phi\left(\mathbb{X}_{\text{g}}\right)$.

Additional metrics of interest are the surface $\phi\left(\overline{\mathbb{X}}_{i,k}^{\text{t}}\right)$
of the set $\overline{\mathbb{X}}_{i,k}^{\text{t}}$, and the surface
$\phi\left(\overline{\mathbb{X}}_{i,k}^{\text{t}}\cap\mathbb{X}_{i,k}^{\text{h}}\right)$
of the part of $\overline{\mathbb{X}}_{i,k}^{\text{t}}$ which has
been occluded by an obstacle. In this last metric, the set $\mathbb{X}_{i,k}^{\text{h}}$
represents the portion of $\mathbb{X}_{\text{g}}$ that has not been
seen by UAV~$i$ at a previous time instant due to an occlusion by
an obstacle and has not yet been observed again up to time $t_{k}$.
This set is initialized with $\mathbb{X}_{i,0}^{\text{h}}=\emptyset$
and updated as
\begin{align}
\mathbb{X}_{i,k\mid k}^{\text{h}} & =\left(\mathbb{X}_{i,k-1}^{\text{h}}\cup\mathbb{H}_{i,k}\right)\setminus\mathbb{P}_{i,k}^{\text{g}}\left(\mathcal{Y}_{i,k}^{\text{g}}\right),\label{eq:Xh}
\end{align}
Moreover, using information received from its neighbors, UAV~$i$
updates $\mathbb{X}_{i,k\mid k}^{\text{h}}$ as
\begin{equation}
\mathbb{X}_{i,k}^{\text{h}}=\left({\textstyle \bigcup}_{\ell\in\mathcal{N}_{i,k}}\mathbb{X}_{\ell,k\mid k}^{\text{h}}\right)\setminus\left({\textstyle \bigcup}_{\ell\in\mathcal{N}_{i,k}}\mathbb{P}_{\ell,k}^{\text{g}}\left(\mathcal{Y}_{\ell,k}^{\text{g}}\right)\right).
\end{equation}

\subsection{Results}

Results have been averaged over $10$ independent simulations, each
lasting $300$~s.A video illustrating the evaluation of the set estimates
for one simulation is available at: https://nextcloud.centralesupelec.fr/s/Sxf5Nk2RJRXMyzk.

\begin{figure}
\begin{centering}
\includegraphics[width=0.8\columnwidth]{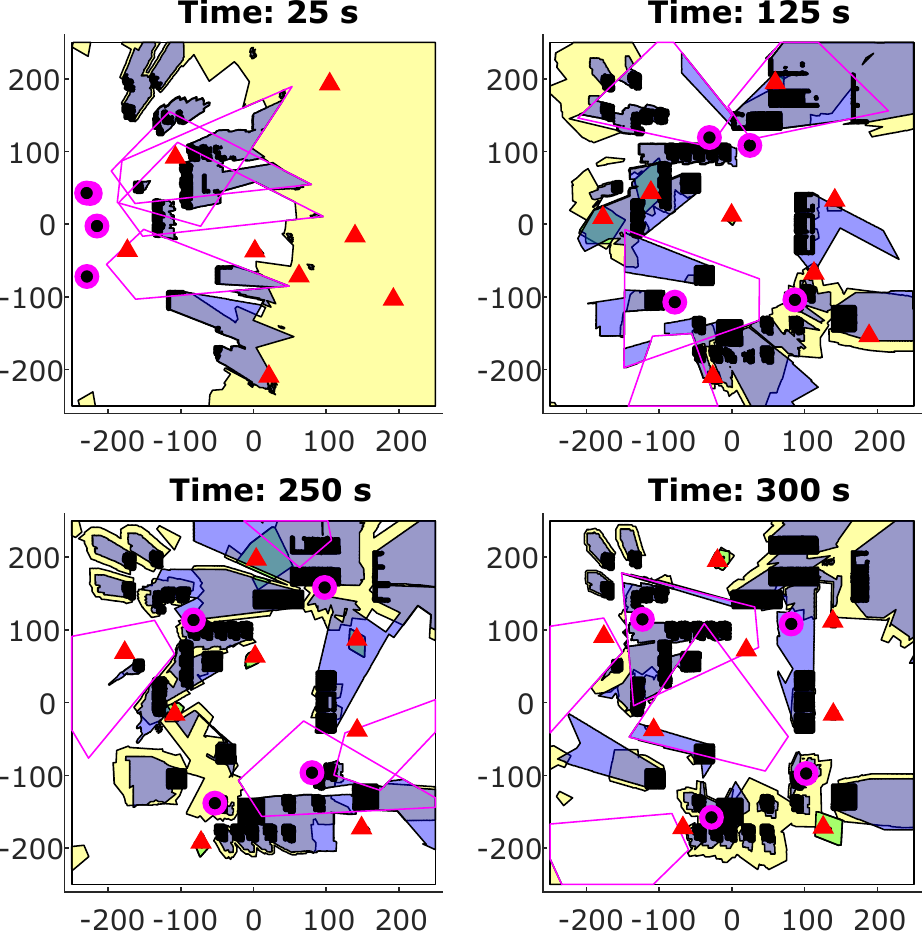}
\par\end{centering}
\caption{\label{fig:Simu_SetEstimateEvolution} Set estimates $\mathbb{X}_{i,j,k}^{\text{t}}$
(green), $\overline{\mathbb{X}}_{i,k}^{\text{t}}$ (yellow), $\mathbb{X}_{i,k}^{\text{o}}$
(black), and $\mathbb{X}_{i,k}^{\text{h}}$ (blue) at different time
instant; The UAVs and the intersection of their FoV with the ground
are represented in purple; Targets are represented by red triangles.}
\end{figure}

Figure~\ref{fig:Simu_SetEstimateEvolution} shows the sets $\overline{\mathbb{X}}_{i,k}^{\text{t}}$
(in yellow), $\mathbb{X}_{i,j,k}^{\text{t}}$ (in green), $\mathbb{X}_{i,k}^{\text{o}}$
(in black), and $\mathbb{X}_{i,k}^{\text{h}}$ (in blue) for a typical
realization at $4$ time instants. Since there is no restriction on
their communication range, all UAVs share the same set estimates.

\begin{figure}
\begin{centering}
\includegraphics[width=0.8\columnwidth]{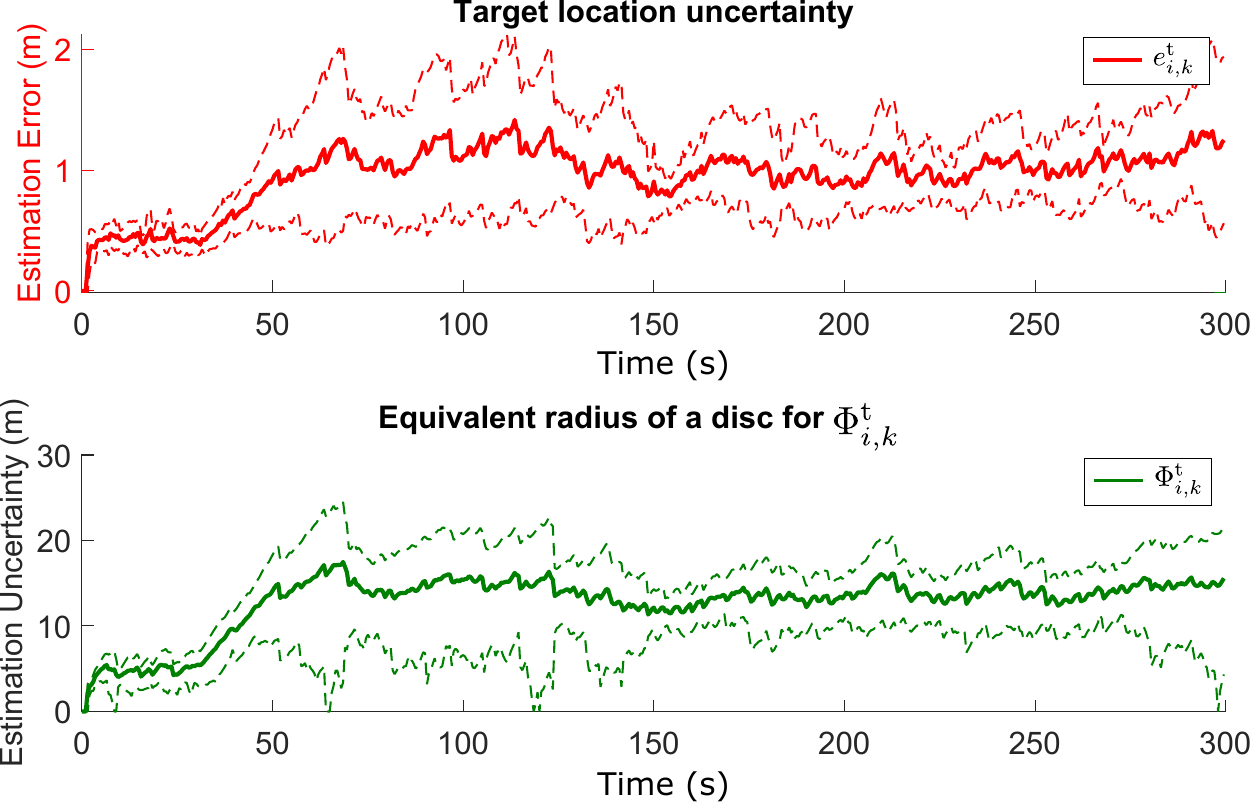}
\par\end{centering}
\caption{\label{fig:Simu_1}Mean value and standard deviation over 10 simulations
of $e_{i,k}^{\text{t}}$ (top) and of the radius of a disc of the
same area as $\Phi_{i,k}^{\text{t}}$ (bottom).}
\end{figure}

Figure~\ref{fig:Simu_1} (top) illustrates the evolution of $e_{i,k}^{\text{t}}$
(in red). Between $t=150$~s and $t=300$~s, the average target
localization error $e_{i,k}^{\text{t}}$ evolves around $1.0\pm0.4\,\text{m}$.
Figure~\ref{fig:Simu_1} (bottom) shows the evolution of the radius
of a disc which area is $\Phi_{i,k}^{\text{t}}$. Between $t=150$~s
and $t=300$~s, it is around $14\pm5$~m. This localization error
remains small, but the uncertainty is relatively large. This is due
to the fact that there are not enough UAVs to permanently observe
the targets. In absence of observations, the uncertainty grows. Moreover,
the absence of knowledge of the dynamics of targets (apart from their
maximal velocity), does not enable UAVs to efficiently predict the
evolution of the identified targets.

\begin{figure}
\begin{centering}
\includegraphics[width=0.8\columnwidth]{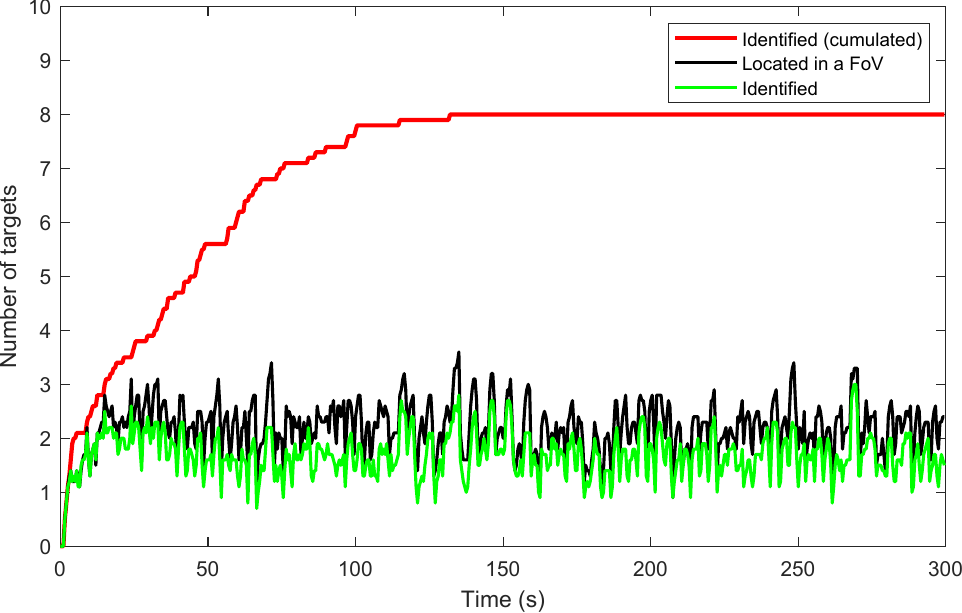}
\par\end{centering}
\caption{\label{fig:Simu_2}Evolution with time of $\text{card}\left(\mathcal{L}_{i,k}^{\text{t}}\right)$
(red), of the number of targets located in the FoV of at least one
UAV (black), and of the number of identified targets, \emph{i.e.},
of $\text{card}\left(\bigcup_{\ell\in\mathcal{N}^{\text{u}}}\mathcal{D}_{\ell,k}^{\text{t}}\right)$
(green).}
\end{figure}

Figure~\ref{fig:Simu_2} shows $\text{card}\left(\mathcal{L}_{i,k}^{\text{t}}\right)$,
the cumulated number of detected targets (in red). The UAVs are able
to identify all targets in less than $150$~seconds. Figure~\ref{fig:Simu_2}
also shows the number of targets included in the FoV of all UAVs (in
black) and the number of targets that are identified, \emph{i.e.},
$\text{card}\left(\bigcup_{\ell\in\mathcal{N}^{\text{u}}}\mathcal{D}_{\ell,k}^{\text{t}}\right)$
(in blue). The difference comes from the obstacles hiding some targets,
or targets located in the FoV, but too far away to be properly identified.

\begin{figure}
\begin{centering}
\includegraphics[width=0.8\columnwidth]{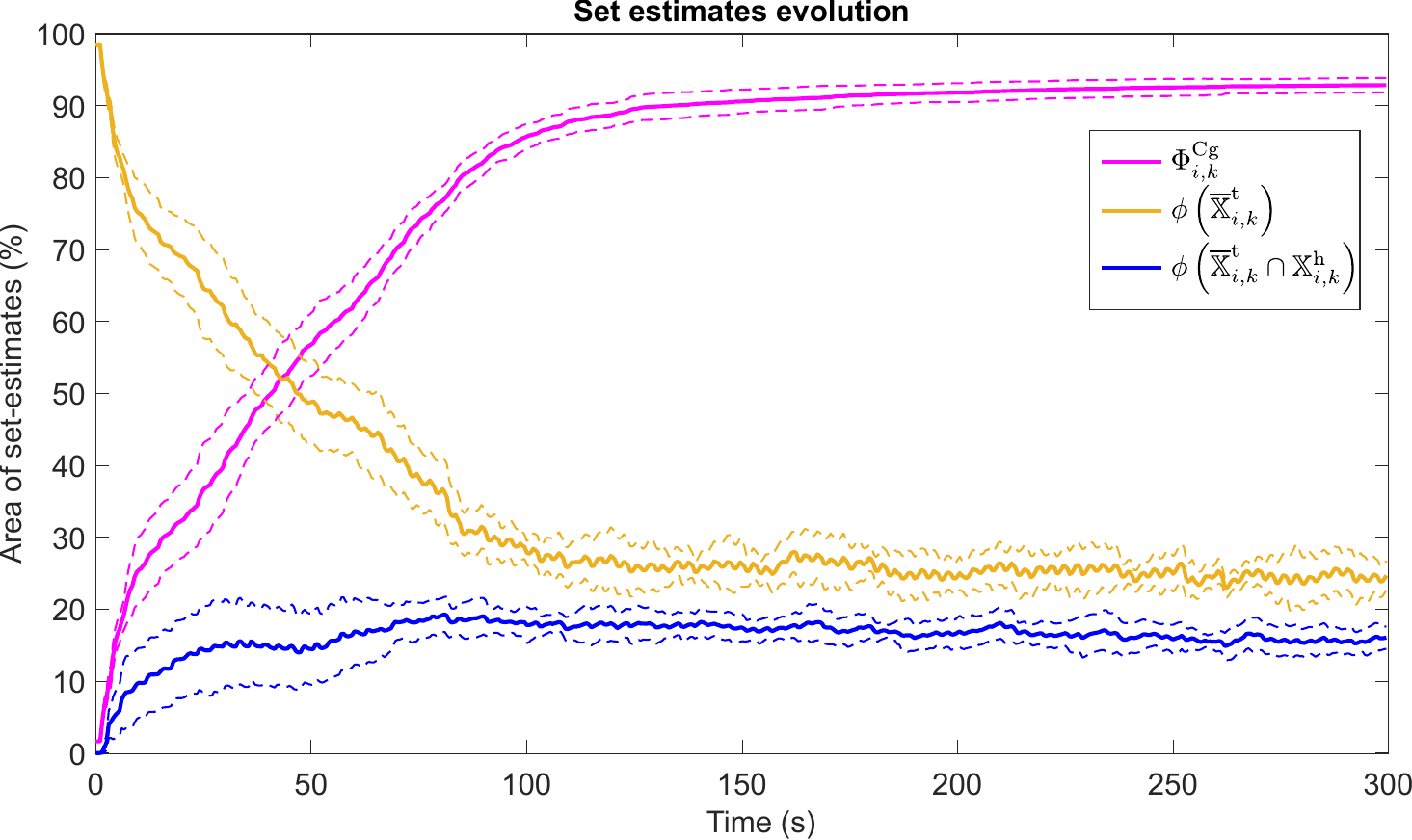}
\par\end{centering}
\caption{\label{fig:Simu_3}Top: mean value with standard deviation over 10
simulations of $\Phi_{i,k}^{\text{Cg}}$ (purple), $\phi\left(\overline{\mathbb{X}}_{i,k}^{\text{t}}\right)/\phi\left(\mathbb{X}_{\text{g}}\right)$
(orange), $\phi\left(\overline{\mathbb{X}}_{i,k}^{\text{t}}\cap\mathbb{X}_{i,k}^{\text{h}}\right)/\phi\left(\mathbb{X}_{\text{g}}\right)$
(blue).}
\end{figure}

Figure~\ref{fig:Simu_3} shows the evolution of $\phi\left(\overline{\mathbb{X}}_{i,k}^{\text{t}}\right)/\phi\left(\mathbb{X}_{\text{g}}\right)$
(in orange), of $\phi\left(\overline{\mathbb{X}}_{i,k}^{\text{t}}\cap\mathbb{X}_{i,k}^{\text{h}}\right)/\phi\left(\mathbb{X}_{\text{g}}\right)$
(in blue) and of $\Phi_{i,k}^{\text{Cg}}/\phi\left(\mathbb{X}_{\text{g}}\right)$
(in purple). One observes that $\phi\left(\overline{\mathbb{X}}_{i,k}^{\text{t}}\right)$
decreases quickly during the first time steps and remains around $26\pm3\,\text{\%}$
of $\phi\left(\mathbb{X}_{\text{g}}\right)$. As $\overline{\mathbb{X}}_{i,k}^{\text{t}}$
does not become empty, the fleet is unable to ensure that all targets
have been found. The evolution of $\phi\left(\overline{\mathbb{X}}_{i,k}^{\text{t}}\cap\mathbb{X}_{i,k}^{\text{h}}\right)/\phi\left(\mathbb{X}_{\text{g}}\right)$
shows that after $t=150\,\text{s}$, almost two third of $\overline{\mathbb{X}}_{i,k}^{\text{t}}$
belongs to $\overline{\mathbb{X}}_{i,k}^{\text{t}}\cap\mathbb{X}_{i,k}^{\text{h}}$,
and to parts of $\mathbb{X}_{\text{g}}$ that have been hidden by
obstacles. These parts of $\overline{\mathbb{X}}_{i,k}^{\text{t}}$
intersecting $\mathbb{X}_{i,k}^{\text{h}}$ grow until they are observed
from a different point of view. This is clearly a limitation of the
trajectory design approach, which is unable to determine the best
point of view to observe some previously hidden areas.

$\Phi_{i,k}^{\text{Cg}}$ increases quickly at the beginning until
all targets have been identified. After $150$~s, $\Phi_{i,k}^{\text{Cg}}/\phi\left(\mathbb{X}_{\text{g}}\right)$
is above $90$~\% and reaches about $93$~\% at the end of the simulations.
Since $5$\% of $\mathbb{X}_{\text{g}}$ is occupied by obstacles,
the maximum theoretical value of $\Phi_{i,k}^{\text{Cg}}/\phi\left(\mathbb{X}_{\text{g}}\right)$
is $95$\%. UAVs have explored nearly all the RoI. The remaining $2$~\%
correspond to areas around buildings located close to the borders
of the RoI. This is a second issue of the trajectory design approach:
the criterion~(\ref{eq:Crit}) favors reduction of sets close to
UAVs and does not impose the full exploration of the RoI.

\begin{figure}
\begin{centering}
\includegraphics[width=0.8\columnwidth]{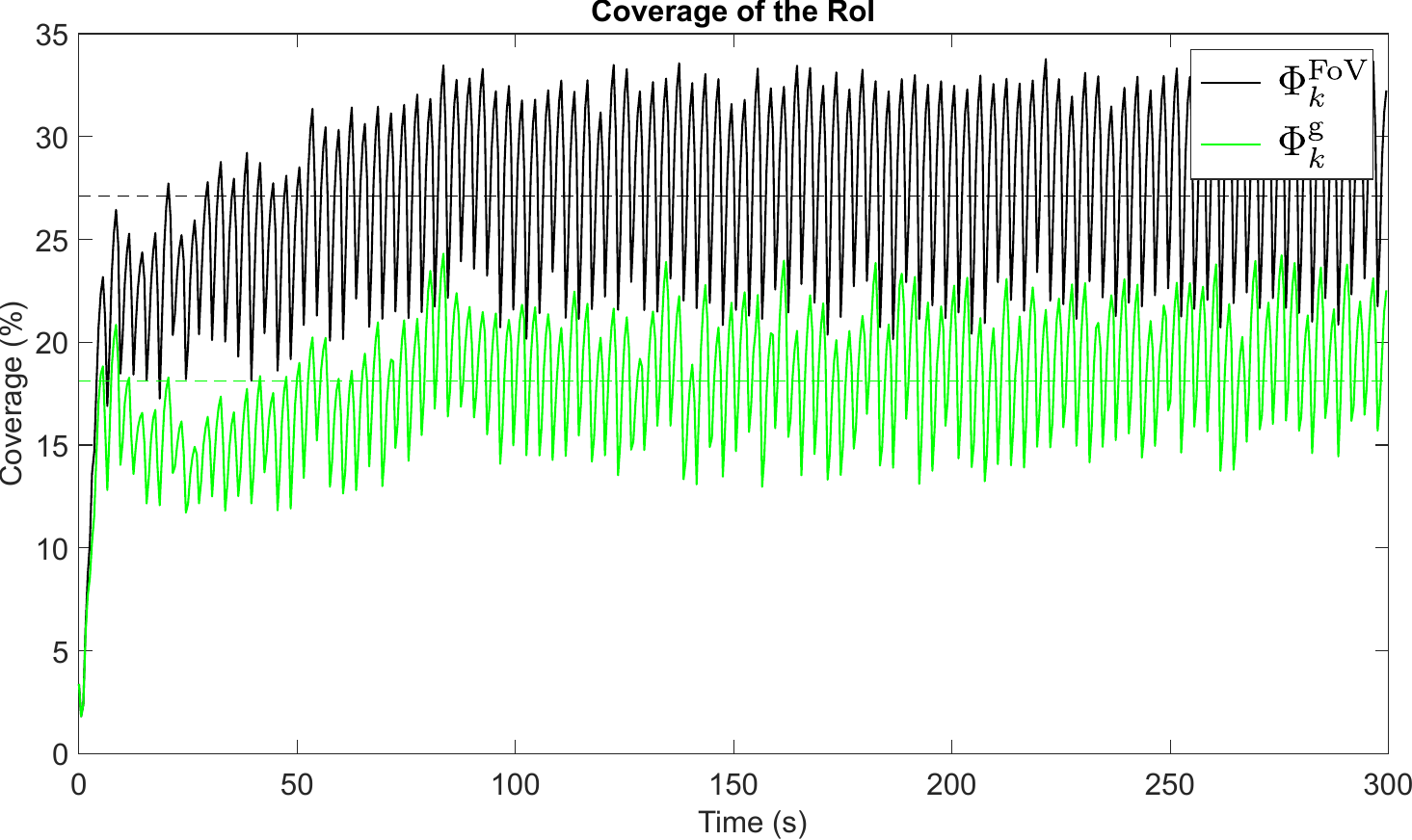}
\par\end{centering}
\caption{\label{fig:Coverage-1}Mean value of $\Phi_{k}^{\text{FoV}}/\phi\left(\mathbb{X}_{\text{g}}\right)$
(black) and of $\Phi_{k}^{\text{g}}/\phi\left(\mathbb{X}_{\text{g}}\right)$
(green); the dashed lines represent the corresponding average over
time.}
\end{figure}

Figure~\ref{fig:Coverage-1} illustrates the evolution of $\Phi_{k}^{\text{FoV}}/\phi\left(\mathbb{X}_{\text{g}}\right)$
and of $\Phi_{k}^{\text{g}}/\phi\left(\mathbb{X}_{\text{g}}\right)$.
Initially, both quantities are small, as the UAVs are launched from
the same location. Their FoVs intersect significantly. After few time
steps, the UAVs are able to spread and at each time instant to monitor
about $27$\% of the RoI. Due to the presence of obstacles, the effective
ground area observed by the fleet reduces to about $18$\%. The oscillations
of both curves are due to the update of the trajectories with a period
of $3$~s.

\section{Conclusion\label{sec:Conclusion}}

This paper presents an approach to solve a CSAT problem in an unknown
and cluttered environment with set-membership approach exploiting
CVS information. For that purpose, several hypotheses related to the
information provided by CVS are introduced to make such information
exploitable by set-membership techniques. They consist mainly of a
model for a depth measurements, and a model for the interpretation
of the pixel classification. 

Based on these hypotheses, CVS information is exploited by the proposed
set-membership algorithm in each UAV to characterize sets guaranteed
to contain the true location of identified targets and a set potentially
containing the targets remaining to detect. The remainder of the RoI
is proved to be free of targets. Via communications, the set-membership
approach allows an easy exploitation of estimates provided the neighbors
of each UAV to get refined estimates. Finally, each UAV updates its
trajectory using a cooperative MPC approach.

Simulation results have shown the performance of the proposed set-membership
CSAT algorithm considering a simplified urban environment. Some limitations
have also been evidenced regarding the ability of the fleet to entirely
explore the RoI and to observe area hidden by obstacles with the best
point of view. These issues will be addressed in future work by building
a 2.5D or a 3D map \cite{souza_occupancy-elevation_2016} of the environment
during exploration. This will allow UAVs to predict the portions of
the ground occluded by obstacles during their trajectory design.
Other research directions may include the processing of decoys erroneously
detected as targets by the CVS.


\begin{thebibliography}{10}

\bibitem{robin_multi-robot_2016}
C.~Robin and S.~Lacroix, ``Multi-robot target detection and tracking: taxonomy
  and survey,'' {\em Autonomous Robots}, vol.~40, no.~4, pp.~729--760, 2016.

\bibitem{queralta_collaborative_2020}
J.~P. Queralta, J.~Taipalmaa, B.~C. Pullinen, V.~K. Sarker, T.~N. Gia,
  H.~Tenhunen, M.~Gabbouj, J.~Raitoharju, and T.~Westerlund, ``Collaborative
  multi-robot search and rescue: {Planning}, coordination, perception, and
  active vision,'' {\em IEEE Access}, vol.~8, pp.~191617--191643, 2020.

\bibitem{abood_survey_2022}
S.~S. Abood, K.~Q. Hussein, and M.~T. Gaata, ``Survey on {Modern}
  {Applications} of {Multiple} {Unmanned} {Aerial} {Vehicles} ({UAV})
  {Systems},'' in {\em Proc. {IEEE} {CSCTIT}}, pp.~179--184, IEEE, 2022.

\bibitem{lyu_unmanned_2023}
M.~Lyu, Y.~Zhao, C.~Huang, and H.~Huang, ``Unmanned aerial vehicles for search
  and rescue: {A} survey,'' {\em Remote Sensing}, vol.~15, no.~13, p.~3266,
  2023.

\bibitem{li_target_2021}
J.~Li, X.~Zhai, J.~Xu, and C.~Li, ``Target search algorithm for {AUV} based on
  real-time perception maps in unknown environment,'' {\em Machines}, vol.~9,
  no.~8, pp.~147--173, 2021.

\bibitem{ji_source_2022}
Y.~Ji, Y.~Zhao, B.~Chen, Z.~Zhu, Y.~Liu, H.~Zhu, and S.~Qiu, ``Source searching
  in unknown obstructed environments through source estimation, target
  determination, and path planning,'' {\em Building and Environment}, vol.~221,
  pp.~109266--109306, 2022.

\bibitem{placed_survey_2023}
J.~A. Placed, J.~Strader, H.~Carrillo, N.~Atanasov, V.~Indelman, L.~Carlone,
  and J.~A. Castellanos, ``A survey on active simultaneous localization and
  mapping: {State} of the art and new frontiers,'' {\em IEEE Trans. Robotics},
  vol.~39, no.~3, pp.~1686--1705, 2023.

\bibitem{kuhlman_multipass_2017}
M.~J. Kuhlman, M.~W. Otte, D.~Sofge, and S.~K. Gupta, ``Multipass target search
  in natural environments,'' {\em Sensors}, vol.~17, no.~11, pp.~2514--2550,
  2017.

\bibitem{zhu_multi-uav_2021}
X.~Zhu, F.~Vanegas, F.~Gonzalez, and C.~Sanderson, ``A multi-{UAV} system for
  exploration and target finding in cluttered and {GPS}-denied environments,''
  in {\em Proc. {IEEE} {ICUAS}}, pp.~721--729, 2021.

\bibitem{zhao_distributed_2022}
L.~Zhao, R.~Li, J.~Han, and J.~Zhang, ``A distributed model predictive
  control-based method for multidifferent-target search in unknown
  environments,'' {\em IEEE Trans. Evol. Comput.}, vol.~27, no.~1,
  pp.~111--125, 2022.

\bibitem{ibenthal_localization_2023}
J.~Ibenthal, L.~Meyer, H.~Piet-Lahanier, and M.~Kieffer, ``Localization of
  {Partially} {Hidden} {Moving} {Targets} {Using} a {Fleet} of {UAVs} via
  {Bounded}-{Error} {Estimation},'' {\em IEEE Trans. Robotics}, vol.~39, no.~6,
  pp.~4211--4229, 2023.

\bibitem{zhang_enhancing_2024}
B.~Zhang, X.~Lin, Y.~Zhu, J.~Tian, and Z.~Zhu, ``Enhancing {Multi}-{UAV}
  {Reconnaissance} and {Search} {Through} {Double} {Critic} {DDPG} {With}
  {Belief} {Probability} {Maps},'' {\em IEEE Trans. Intelligent Vehicles},
  vol.~9, no.~2, pp.~3827--3842, 2024.

\bibitem{ibenthal_bounded-error_2021}
J.~Ibenthal, M.~Kieffer, L.~Meyer, H.~Piet-Lahanier, and S.~Reynaud,
  ``Bounded-error target localization and tracking using a fleet of {UAVs},''
  {\em Automatica}, vol.~132, pp.~109809--109824, 2021.

\bibitem{dames_distributed_2020}
P.~M. Dames, ``Distributed multi-target search and tracking using the {PHD}
  filter,'' {\em Autonomous robots}, vol.~44, no.~3, pp.~673--689, 2020.

\bibitem{hou_uav_2023}
Y.~Hou, J.~Zhao, R.~Zhang, X.~Cheng, and L.~Yang, ``{UAV} {Swarm} {Cooperative}
  {Target} {Search}: {A} {Multi}-{Agent} {Reinforcement} {Learning}
  {Approach},'' {\em IEEE Trans. Intelligent Vehicles}, vol.~9, no.~1,
  pp.~568--578, 2023.

\bibitem{banerjee_decentralized_2024}
A.~Banerjee and J.~Schneider, ``Decentralized {Multi}-{Agent} {Active} {Search}
  and {Tracking} when {Targets} {Outnumber} {Agents},'' {\em arXiv preprint
  arXiv:2401.03154}, 2024.

\bibitem{tang_novel_2019}
H.~Tang, W.~Sun, H.~Yu, A.~Lin, M.~Xue, and Y.~Song, ``A novel hybrid algorithm
  based on {PSO} and {FOA} for target searching in unknown environments,'' {\em
  Applied Intelligence}, vol.~49, pp.~2603--2622, 2019.

\bibitem{vanegas_uav_2016}
F.~Vanegas, D.~Campbell, M.~Eich, and F.~Gonzalez, ``{UAV} based target finding
  and tracking in {GPS}-denied and cluttered environments,'' in {\em Proc.
  {IEEE} {IROS}}, pp.~2307--2313, 2016.

\bibitem{goldhoorn_searching_2018}
A.~Goldhoorn, A.~Garrell, R.~Alqu�zar, and A.~Sanfeliu, ``Searching and
  tracking people with cooperative mobile robots,'' {\em Autonomous Robots},
  vol.~42, no.~4, pp.~739--759, 2018.

\bibitem{meera_obstacle-aware_2019}
A.~A. Meera, M.~Popovic, A.~Millane, and R.~Siegwart, ``Obstacle-aware adaptive
  informative path planning for uav-based target search,'' in {\em Proc. {IEEE}
  {ICRA}}, pp.~718--724, 2019.

\bibitem{reboul_cooperative_2019}
L.~Reboul, M.~Kieffer, H.~Piet-Lahanier, and S.~Reynaud, ``Cooperative guidance
  of a fleet of {UAVs} for multi-target discovery and tracking in presence of
  obstacles using a set membership approach,'' {\em IFAC-PapersOnLine},
  vol.~52, no.~12, pp.~340--345, 2019.

\bibitem{hardouin_next-best-view_2020}
G.~Hardouin, J.~Moras, F.~Morbidi, J.~Marzat, and E.~M. Mouaddib,
  ``Next-{Best}-{View} planning for surface reconstruction of large-scale {3D}
  environments with multiple {UAVs},'' in {\em Proc. {IEEE} {IROS}},
  pp.~1567--1574, 2020.

\bibitem{asgharivaskasi_semantic_2023}
A.~Asgharivaskasi and N.~Atanasov, ``Semantic {OcTree} mapping and {Shannon}
  mutual information computation for robot exploration,'' {\em IEEE Trans.
  Robotics}, vol.~39, no.~3, pp.~1910--1928, 2023.

\bibitem{zhang_vehicle_2023}
H.~Zhang, C.~Xie, H.~Toriya, H.~Shishido, and I.~Kitahara, ``Vehicle
  {Localization} in a {Completed} {City}-{Scale} {3D} {Scene} {Using} {Aerial}
  {Images} and an {On}-{Board} {Stereo} {Camera},'' {\em Remote Sensing},
  vol.~15, no.~15, pp.~3871--3891, 2023.

\bibitem{atanasov_nonmyopic_2014}
N.~Atanasov, B.~Sankaran, J.~Le~Ny, G.~J. Pappas, and K.~Daniilidis,
  ``Nonmyopic view planning for active object classification and pose
  estimation,'' {\em IEEE Trans. Robotics}, vol.~30, no.~5, pp.~1078--1090,
  2014.

\bibitem{zhong_detect-slam_2018}
F.~Zhong, S.~Wang, Z.~Zhang, and Y.~Wang, ``Detect-{SLAM}: {Making} object
  detection and {SLAM} mutually beneficial,'' in {\em {IEEE} {WACV}},
  pp.~1001--1010, 2018.

\bibitem{zhang_fast_2022}
S.~Zhang, X.~Zhang, T.~Li, J.~Yuan, and Y.~Fang, ``Fast active aerial
  exploration for traversable path finding of ground robots in unknown
  environments,'' {\em IEEE Trans. Instrum. Meas.}, vol.~71, pp.~1--13, 2022.

\bibitem{serdel_smana_2023}
Q.~Serdel, J.~Marzat, and J.~Moras, ``{SMaNa}: {Semantic} {Mapping} and
  {Navigation} {Architecture} for {Autonomous} {Robots},'' {\em Proc. ICINCO},
  vol.~1, pp.~453--464, 2023.

\bibitem{park_stereo_2012}
J.~Park and Y.~Kim, ``Stereo vision based collision avoidance of quadrotor
  {UAV},'' in {\em Proc. {ICCAS}}, pp.~173--178, 2012.

\bibitem{jeon_accurate_2015}
H.-G. Jeon, J.~Park, G.~Choe, J.~Park, Y.~Bok, Y.-W. Tai, and I.~So~Kweon,
  ``Accurate depth map estimation from a lenslet light field camera,'' in {\em
  Proc. {IEEE} {CVPR}}, pp.~1547--1555, 2015.

\bibitem{madhuanand_self-supervised_2021}
L.~Madhuanand, F.~Nex, and M.~Y. Yang, ``Self-supervised monocular depth
  estimation from oblique {UAV} videos,'' {\em ISPRS journal of photogrammetry
  and remote sensing}, vol.~176, pp.~1--14, 2021.

\bibitem{shimada_fast_2023}
T.~Shimada, H.~Nishikawa, X.~Kong, and H.~Tomiyama, ``Fast and {High}-{Quality}
  {Monocular} {Depth} {Estimation} with {Optical} {Flow} for {Autonomous}
  {Drones},'' {\em Drones}, vol.~7, no.~2, pp.~134--151, 2023.

\bibitem{grilli_review_2017}
E.~Grilli, F.~Menna, and F.~Remondino, ``A review of point clouds segmentation
  and classification algorithms,'' {\em ISPRS Archives}, vol.~42, p.~339, 2017.

\bibitem{howard_searching_2019}
A.~Howard, M.~Sandler, G.~Chu, L.-C. Chen, B.~Chen, M.~Tan, W.~Wang, Y.~Zhu,
  R.~Pang, V.~Vasudevan, and {others}, ``Searching for mobilenetv3,'' in {\em
  Proc. {IEEE}/{CVF} {ICCV}}, pp.~1314--1324, 2019.

\bibitem{minaee_image_2021}
S.~Minaee, Y.~Boykov, F.~Porikli, A.~Plaza, N.~Kehtarnavaz, and D.~Terzopoulos,
  ``Image segmentation using deep learning: {A} survey,'' {\em IEEE Trans.
  Pattern Anal. Mach. Intell.}, vol.~44, no.~7, pp.~3523--3542, 2021.

\bibitem{redmon_you_2016}
J.~Redmon, S.~Divvala, R.~Girshick, and A.~Farhadi, ``You only look once:
  {Unified}, real-time object detection,'' in {\em Proc. {IEEE} {CVPR}},
  pp.~779--788, 2016.

\bibitem{minaeian_effective_2018}
S.~Minaeian, J.~Liu, and Y.-J. Son, ``Effective and efficient detection of
  moving targets from a {UAV} camera,'' {\em IEEE Trans. Intell. Transp.
  Syst.}, vol.~19, no.~2, pp.~497--506, 2018.

\bibitem{luo_yolod_2022}
X.~Luo, Y.~Wu, and L.~Zhao, ``{YOLOD}: {A} target detection method for {UAV}
  aerial imagery,'' {\em Remote Sensing}, vol.~14, no.~14, p.~3240, 2022.

\bibitem{jiang_review_2022}
P.~Jiang, D.~Ergu, F.~Liu, Y.~Cai, and B.~Ma, ``A {Review} of {Yolo} algorithm
  developments,'' {\em Procedia computer science}, vol.~199, pp.~1066--1073,
  2022.

\bibitem{minaeian_vision-based_2015}
S.~Minaeian, J.~Liu, and Y.-J. Son, ``Vision-based target detection and
  localization via a team of cooperative {UAV} and {UGVs},'' {\em IEEE Trans.
  Syst. Man. Cybern.}, vol.~46, no.~7, pp.~1005--1016, 2015.

\bibitem{sun_camera-based_2016}
J.~Sun, B.~Li, Y.~Jiang, and C.-y. Wen, ``A camera-based target detection and
  positioning {UAV} system for search and rescue ({SAR}) purposes,'' {\em
  Sensors}, vol.~16, no.~11, pp.~1778--1802, 2016.

\bibitem{wang_real-time_2016}
X.~Wang, J.~Liu, and Q.~Zhou, ``Real-time multi-target localization from
  unmanned aerial vehicles,'' {\em Sensors}, vol.~17, no.~1, pp.~33--61, 2016.

\bibitem{liu_novel_2018}
Y.~Liu, Q.~Wang, H.~Hu, and Y.~He, ``A novel real-time moving target tracking
  and path planning system for a quadrotor {UAV} in unknown unstructured
  outdoor scenes,'' {\em IEEE Trans. Syst. Man. Cybern.}, vol.~49, no.~11,
  pp.~2362--2372, 2018.

\bibitem{liu_vision-aware_2022}
D.~Liu, W.~Bao, X.~Zhu, B.~Fei, Z.~Xiao, and T.~Men, ``Vision-aware air-ground
  cooperative target localization for {UAV} and {UGV},'' {\em Aerosp. Sci.
  Technol.}, vol.~124, pp.~107525--107540, 2022.

\bibitem{di_gennaro_sensor_2023}
T.~M. Di~Gennaro and J.~Waldmann, ``Sensor {Fusion} with {Asynchronous}
  {Decentralized} {Processing} for {3D} {Target} {Tracking} with a {Wireless}
  {Camera} {Network},'' {\em Sensors}, vol.~23, no.~3, pp.~1194--1228, 2023.

\bibitem{kenmogne_cooperative_2019}
I.-F. Kenmogne, V.~Drevelle, and E.~Marchand, ``Cooperative localization of
  drones by using interval methods,'' {\em Acta Cybernetica}, pp.~1--16, 2019.

\bibitem{ajmera_autonomous_2020}
Y.~Ajmera and S.~P. Singh, ``Autonomous {UAV}-based {Target} {Search},
  {Tracking} and {Following} using {Reinforcement} {Learning} and {YOLOFlow},''
  in {\em {IEEE} {SSRR}}, pp.~15--20, 2020.

\bibitem{liu_target_2022}
Y.~Liu, M.~Xu, G.~Jiang, X.~Tong, J.~Yun, Y.~Liu, B.~Chen, Y.~Cao, N.~Sun, and
  Z.~Li, ``Target localization in local dense mapping using {RGBD} {SLAM} and
  object detection,'' {\em Concurrency and Computation: Practice and
  Experience}, vol.~34, no.~4, p.~e6655, 2022.

\bibitem{zeng_semantic_2018}
Z.~Zeng, Y.~Zhou, O.~C. Jenkins, and K.~Desingh, ``Semantic mapping with
  simultaneous object detection and localization,'' in {\em Proc. {IEEE}
  {IROS}}, pp.~911--918, 2018.

\bibitem{sahin_review_2020}
C.~Sahin, G.~Garcia-Hernando, J.~Sock, and T.-K. Kim, ``A review on object pose
  recovery: from 3d bounding box detectors to full 6d pose estimators,'' {\em
  Image. Vision. Comput.}, vol.~96, p.~103898, 2020.

\bibitem{fan_deep_2022}
Z.~Fan, Y.~Zhu, Y.~He, Q.~Sun, H.~Liu, and J.~He, ``Deep learning on monocular
  object pose detection and tracking: {A} comprehensive overview,'' {\em ACM
  Computing Surveys}, vol.~55, no.~4, pp.~1--40, 2022.

\bibitem{mousavian_3d_2017}
A.~Mousavian, D.~Anguelov, J.~Flynn, and J.~Kosecka, ``3d bounding box
  estimation using deep learning and geometry,'' in {\em Pro. {IEEE} {CVPR}},
  pp.~7074--7082, 2017.

\bibitem{zhen_intelligent_2020}
Z.~Zhen, Y.~Chen, L.~Wen, and B.~Han, ``An intelligent cooperative mission
  planning scheme of {UAV} swarm in uncertain dynamic environment,'' {\em
  Aerospace Science and Technology}, vol.~100, pp.~105826--105842, 2020.

\bibitem{allik_tracking_2019}
B.~Allik, ``Tracking of multiple targets across distributed platforms with fov
  constraints,'' in {\em Proc. {IEEE} {CDC}}, pp.~6044--6049, 2019.

\bibitem{symington_probabilistic_2010}
A.~Symington, S.~Waharte, S.~Julier, and N.~Trigoni, ``Probabilistic target
  detection by camera-equipped {UAVs},'' in {\em Proc. {IEEE} {ICRA}},
  pp.~4076--4081, 2010.

\bibitem{niedzielski_first_2021}
T.~Niedzielski, M.~Jurecka, B.~Mizinski, W.~Pawul, and T.~Motyl, ``First
  successful rescue of a lost person using the human detection system: {A} case
  study from {Beskid} {Niski} ({SE} {Poland}),'' {\em Remote Sensing}, vol.~13,
  no.~23, pp.~4903--4921, 2021.

\bibitem{yanmaz_joint_2023}
E.~Yanmaz, ``Joint or decoupled optimization: {Multi}-{UAV} path planning for
  search and rescue,'' {\em Ad Hoc Networks}, vol.~138, pp.~103018--103031,
  2023.

\bibitem{singh_modeling_2010}
A.~Singh, F.~Ramos, H.~D. Whyte, and W.~J. Kaiser, ``Modeling and decision
  making in spatio-temporal processes for environmental surveillance,'' in {\em
  Proc. {IEEE} {ICRA}}, pp.~5490--5497, 2010.

\bibitem{faugeras_three-dimensional_1993}
O.~Faugeras, {\em Three-dimensional computer vision: a geometric viewpoint}.
\newblock MIT press, 1993.

\bibitem{christofides_distributed_2013}
P.~D. Christofides, R.~Scattolini, D.~M. De~La~Pena, and J.~Liu, ``Distributed
  model predictive control: {A} tutorial review and future research
  directions,'' {\em Comput Chem Eng}, vol.~51, pp.~21--41, 2013.

\bibitem{webots_httpwwwcyberboticscom_nodate}
{Webots}, ``http://www.cyberbotics.com.''

\bibitem{souza_occupancy-elevation_2016}
A.~Souza and L.~M. Gon\c calves, ``Occupancy-elevation grid: an alternative
  approach for robotic mapping and navigation,'' {\em Robotica}, vol.~34,
  no.~11, pp.~2592--2609, 2016.

\end{thebibliography}

\appendix

\section{Characterization of the set estimates\label{sec:Set_Characterisation}}

\subsection{Characterization of $\mathbb{X}_{i,j}^{\text{t,m}}$ and $\underline{\mathbb{X}}_{i}^{\text{o}}$\label{subsec:Appendix_CharacX}}

This section describes an approach to get an outer-approximation of
the set estimate $\mathbb{X}_{i,j}^{\text{t,m}}$ introduced in Section~\ref{subsec:setEstimate}
as well as an inner-approximation $\underline{\mathbb{X}}_{i}^{\text{o}}$
of the $r^{\text{s}}$-ground neighborhood of $\bigcup_{m\in\mathcal{N}^{\text{o}}}p_{\text{g}}\left(\mathbb{S}_{m}^{\text{o}}\right)$
introduced in Section~\ref{subsec:FreeSpace_Obs}. Both sets requires
the characterization of $\mathbb{P}_{i}\left(\left(n_{\text{r}},n_{\text{c}}\right)\right)$
defined in (\ref{eq:Pi}).

For any pixel $\left(n_{\text{r}},n_{\text{c}}\right)$, the set $\mathbb{P}_{i}\left(\left(n_{\text{r}},n_{\text{c}}\right)\right)$
is the intersection between (\emph{i}) the half cone of apex $\boldsymbol{x}_{i}^{\text{c}}$
and vertices corresponding to one of the four unit vectors $\mathbf{M}_{\mathcal{F}_{i}^{\text{c}}}^{\mathcal{F}}\boldsymbol{v}_{\ell}^{\mathcal{F}_{i}^{\text{c}}}$,
$\ell=1,\dots,4$ with $\boldsymbol{v}_{1}^{\mathcal{F}_{i}^{\text{c}}}=\boldsymbol{v}^{\mathcal{F}_{i}^{\text{c}}}\left(n_{\text{c}},n_{\text{r}}\right)$,
$\boldsymbol{v}_{2}^{\mathcal{F}_{i}^{\text{c}}}=\boldsymbol{v}^{\mathcal{F}_{i}^{\text{c}}}\left(n_{\text{c}},n_{\text{r}}-1\right)$,
$\boldsymbol{v}_{3}^{\mathcal{F}_{i}^{\text{c}}}=\boldsymbol{v}^{\mathcal{F}_{i}^{\text{c}}}\left(n_{\text{c}}-1,n_{\text{r}}-1\right)$,
and $\boldsymbol{v}_{4}^{\mathcal{F}_{i}^{\text{c}}}=\boldsymbol{v}^{\mathcal{F}_{i}^{\text{c}}}\left(n_{\text{c}},n_{\text{r}}-1\right)$
and (\emph{ii}) the space between the two spheres of center $\boldsymbol{x}_{i}^{\text{c}}$
and radii $\frac{1}{1+\overline{w}}\mathbf{D}_{i}\left(n_{\text{r}},n_{\text{c}}\right)$
and $\frac{1}{1+\underline{w}}\mathbf{D}_{i}\left(n_{\text{r}},n_{\text{c}}\right)$.
Figure~\ref{fig:Pinhole-model} illustrates these four vectors $\boldsymbol{v}_{\ell}^{\mathcal{F}_{i}^{\text{c}}}$
which direction is represented by blue lines.

A truncated pyramidal outer-approximation $\overline{\mathbb{P}}_{i}\left(\left(n_{\text{r}},n_{\text{c}}\right)\right)$
of $\mathbb{P}_{i}\left(\left(n_{\text{r}},n_{\text{c}}\right)\right)$
is obtained considering the angles $\theta_{\ell}$ between $\boldsymbol{v}^{\mathcal{F}_{i}^{\text{c}}}\left(n_{\text{c}}-0.5,n_{\text{r}}-0.5\right)$,
the unit vector supporting the light-ray illuminating the center of
the pixel and $\boldsymbol{v}_{\ell}^{\mathcal{F}_{i}^{\text{c}}}$,
$\ell\in\left\{ 1,2,3,4\right\} $. The eight vertices of $\overline{\mathbb{P}}_{i}\left(\left(n_{\text{r}},n_{\text{c}}\right)\right)$
are then
\begin{align}
\underline{\boldsymbol{x}}_{\ell}^{\text{p}} & \left(n_{\text{r}},n_{\text{c}}\right)=\boldsymbol{x}_{i}^{\text{c}}+\frac{1}{1+\overline{w}}\mathbf{D}_{i}\left(n_{\text{r}},n_{\text{c}}\right)\mathbf{M}_{\mathcal{F}_{i}^{\text{c}}}^{\mathcal{F}}\boldsymbol{v}_{\ell}^{\mathcal{F}_{i}^{\text{c}}},\\
\overline{\boldsymbol{x}}_{\ell}^{\text{p}} & \left(n_{\text{r}},n_{\text{c}}\right)=\boldsymbol{x}_{i}^{\text{c}}+\frac{1}{\cos\left(\theta_{\ell}\right)}\frac{1}{1+\underline{w}}\mathbf{D}_{i}\left(n_{\text{r}},n_{\text{c}}\right)\mathbf{M}_{\mathcal{F}_{i}^{\text{c}}}^{\mathcal{F}}\boldsymbol{v}_{\ell}^{\mathcal{F}_{i}^{\text{c}}},
\end{align}
with $\ell\in\left\{ 1,2,3,4\right\} $.

Then, to evaluate $\mathbb{X}_{i,j}^{\text{t,m}}$ using (\ref{eq:setEstimate_Xijk_bis}),
on determines
\begin{align}
\boldsymbol{p}_{\text{g}}\left(\mathbb{P}_{i,j}^{\text{t}}\right) & ={\textstyle \bigcap}_{\left(n_{\text{r}},n_{\text{c}}\right)\in\left[\mathcal{Y}_{i,j}^{\text{t}}\right]\cap\mathcal{Y}_{i}^{\text{t}}}\boldsymbol{p}_{\text{g}}\left(\mathbb{P}_{i}\left(\left(n_{\text{r}},n_{\text{c}}\right)\right)\right)\\
 & \subset{\textstyle \bigcap}_{\left(n_{\text{r}},n_{\text{c}}\right)\in\left[\mathcal{Y}_{i,j}^{\text{t}}\right]\cap\mathcal{Y}_{i}^{\text{t}}}\boldsymbol{p}_{\text{g}}\left(\overline{\mathbb{P}}_{i}\left(\left(n_{\text{r}},n_{\text{c}}\right)\right)\right),
\end{align}
as $\mathbb{P}_{i}\left(\left(n_{\text{r}},n_{\text{c}}\right)\right)\subset\overline{\mathbb{P}}_{i}\left(\left(n_{\text{r}},n_{\text{c}}\right)\right)$.
Since $\overline{\mathbb{P}}_{i}\left(\left(n_{\text{r}},n_{\text{c}}\right)\right)$
is convex, an outer approximation of $\boldsymbol{p}_{\text{g}}\left(\mathbb{P}_{i}\left(\left(n_{\text{r}},n_{\text{c}}\right)\right)\right)$
is obtained considering the convex hull $\overline{\mathbb{P}}_{\text{g},i}\left(\left(n_{\text{r}},n_{\text{c}}\right)\right)$
of $\mathcal{P}_{\text{g},i}\left(n_{\text{r}},n_{\text{c}}\right)=\left\{ \boldsymbol{p}_{\text{g}}\left(\underline{\boldsymbol{x}}_{\ell}^{\text{p}}\left(n_{\text{r}},n_{\text{c}}\right)\right),\boldsymbol{p}_{\text{g}}\left(\overline{\boldsymbol{x}}_{\ell}^{\text{p}}\left(n_{\text{r}},n_{\text{c}}\right)\right)\right\} _{\ell\in\left\{ 1\dots4\right\} }$.
Finally, an outer-approximation of $\mathbb{X}_{i,j}^{\text{t,m}}$
is obtained as $\mathbb{X}_{\text{g}}\cap\left(\bigcup_{\left(n_{\text{r}},n_{\text{c}}\right)\in\left[\mathcal{Y}_{i,j}^{\text{t}}\right]\cap\mathcal{Y}_{i}^{\text{t}}}\overline{\mathbb{P}}_{\text{g},i}\left(\left(n_{\text{r}},n_{\text{c}}\right)\right)\oplus\boldsymbol{p}_{\text{g}}\left(\mathbb{C}^{\text{t}}\left(\mathbf{0}\right)\right)\right)$.

One has $\underline{\mathbb{X}}_{i}^{\text{o}}=\bigcup_{\left(n_{\text{r}},n_{\text{c}}\right)\in\mathcal{Y}_{i}^{\text{o}}}\mathbb{S}^{\text{o}}\left(\left(n_{\text{r}},n_{\text{c}}\right),r^{\text{s}}\right)$,
see (\ref{eq:Xo_ExclusionSpace}). The definition \ref{eq:So_ExclusionSpace}
can be rewritten as 
\[
\mathbb{S}^{\text{o}}\left(\left(n_{\text{r}},n_{\text{c}}\right),r^{\text{s}}\right)={\textstyle \bigcap}_{\boldsymbol{x}\in\boldsymbol{p}_{\text{g}}\left(\mathbb{P}_{i}\left(\left(n_{\text{r}},n_{\text{c}}\right)\right)\right)}\mathbb{D}_{\text{g}}\left(\boldsymbol{x},r^{\text{s}}\right)
\]
Since $\mathbb{P}_{i}\left(\left(n_{\text{r}},n_{\text{c}}\right)\right)\subset\overline{\mathbb{P}}_{i}\left(\left(n_{\text{r}},n_{\text{c}}\right)\right)$,
one has{\small{}
\begin{equation}
{\textstyle \bigcap}_{\boldsymbol{x}\in\boldsymbol{p}_{\text{g}}\left(\overline{\mathbb{P}}_{i}\left(\left(n_{\text{r}},n_{\text{c}}\right)\right)\right)}\mathbb{D}_{\text{g}}\left(\boldsymbol{x},r^{\text{s}}\right)\subset{\textstyle \bigcap}_{\boldsymbol{x}\in\boldsymbol{p}_{\text{g}}\left(\mathbb{P}_{i}\left(\left(n_{\text{r}},n_{\text{c}}\right)\right)\right)}\mathbb{D}_{\text{g}}\left(\boldsymbol{x},r^{\text{s}}\right)
\end{equation}
}and {\small{}
\begin{equation}
{\textstyle \bigcap}_{\boldsymbol{x}\in\boldsymbol{p}_{\text{g}}\left(\overline{\mathbb{P}}_{i}\left(\left(n_{\text{r}},n_{\text{c}}\right)\right)\right)}\mathbb{D}_{\text{g}}\left(\boldsymbol{x},r^{\text{s}}\right)\subset\mathbb{S}^{\text{o}}\left(\left(n_{r},n_{c}\right),r^{\text{s}}\right).
\end{equation}
}Moreover, as $\overline{\mathbb{P}}_{i}\left(\left(n_{\text{r}},n_{\text{c}}\right)\right)$
is convex,
\[
{\textstyle \bigcap}_{\boldsymbol{x}\in\boldsymbol{p}_{\text{g}}\left(\overline{\mathbb{P}}_{i}\left(\left(n_{\text{r}},n_{\text{c}}\right)\right)\right)}\mathbb{D}_{\text{g}}\left(\boldsymbol{x},r^{\text{s}}\right)={\textstyle \bigcap}{}_{\boldsymbol{x}\in\mathcal{P}_{\text{g},i}\left(n_{\text{r}},n_{\text{c}}\right)}\mathbb{D}_{\text{g}}\left(\boldsymbol{x},r^{\text{s}}\right).
\]
An inner approximation of $\mathbb{S}^{\text{o}}\left(\left(n_{r},n_{c}\right),r_{\text{s}}^{\text{to}}\right)$
is then obtained as the intersection of, at most, eight discs each
with a center in $\mathcal{P}_{\text{g},i}\left(n_{\text{r}},n_{\text{c}}\right)$
and of radius $r^{\text{s}}$. A convex polygon forming an inner-approximation
of this intersection is then easily obtained. The set $\underline{\mathbb{X}}_{i}^{\text{o}}$
is then the union of the polygons obtained for each $\left(n_{\text{r}},n_{\text{c}}\right)\in\mathcal{Y}_{i}^{\text{o}}$.

\subsection{Characterization of $\mathbb{P}_{i}^{\text{g}}\left(\mathcal{Y}_{i}^{\text{g}}\right)$
and $\mathbb{H}_{i}^{\text{g}}$\label{subsec:Appendix_CharacP}}

To evaluate $\mathbb{P}_{i}^{\text{g}}\left(\mathcal{Y}_{i}^{\text{g}}\right)$
introduced in (\ref{eq:Pig}), the union of sets $\mathbb{P}_{i}^{\text{g}}\left(\left(n_{\text{r}},n_{\text{c}}\right)\right)$
for $\left(n_{\text{r}},n_{\text{c}}\right)\in\mathcal{Y}_{i}^{\text{g}}$
has to be characterized. Neglecting the limitation of the FoV by $\mathbb{B}\left(\boldsymbol{x}_{i}^{\text{u}},d_{\max}\right)$
and as $\mathbb{X}_{\text{g}}$ is a part of a plane, $\mathbb{P}_{i}^{\text{g}}\left(\left(n_{\text{r}},n_{\text{c}}\right)\right)$
is the convex quadrangle defined by the intersection of $\mathbb{X}_{\text{g}}$
with the half-cone of apex $\boldsymbol{x}_{i}^{\text{c}}$ and edges
related to the four unit vector $\boldsymbol{v}_{\ell}^{\mathcal{F}_{i}^{\text{c}}}$,
with $\ell=1,\dots,4$, see Figure~\ref{fig:Pinhole-model}. $\mathbb{P}_{i}^{\text{g}}\left(\mathcal{Y}_{i}^{\text{g}}\right)$
is then the union of the convex quadrangles $\mathbb{P}_{i}^{\text{g}}\left(\left(n_{\text{r}},n_{\text{c}}\right)\right)$
for all $\left(n_{\text{r}},n_{\text{c}}\right)\in\mathcal{Y}_{i}^{\text{g}}$.

Similarly, $\mathbb{H}_{i}^{\text{g}}$ is the union of the convex
quadrangles $\mathbb{P}_{i}^{\text{g}}\left(\left(n_{\text{r}},n_{\text{c}}\right)\right)$
for all $\left(n_{\text{r}},n_{\text{c}}\right)\in\mathcal{Y}_{i}^{\text{o}}\cup\mathcal{Y}_{i}^{\text{t}}\cup\mathcal{Y}_{i}^{\text{n}}.$
\end{document}